\documentclass[a4paper,11pt]{article}

\pdfoutput=0 

\usepackage{jinstpub} 

\usepackage{url} 
\usepackage{here}
\usepackage{comment}

\usepackage[driverfallback=dvips]{hyperref} 
\usepackage{mediabb}
\hypersetup{
colorlinks=true,urlcolor=blue,anchorcolor=blue,citecolor=blue,filecolor=blue,linkcolor=blue,menucolor=blue,pagecolor=blue,linktocpage=true,pdfa=true}

\newcommand{\Tabref}[1]{Table~\ref{#1}}
\newcommand{\Equref}[1]{Equation~(\ref{#1})}
\newcommand{\Figref}[1]{Figure~\ref{#1}}
\newcommand*{\mhyph}{\text{-}}

\usepackage{lineno}

\title{\boldmath Application of CNN to a fine segmented scintillator detector for a single particle and neutrino-nucleon event}

\author[a]{Tomohisa Ogawa}

\affiliation[a]{High Energy Accelerator Research Organization,\\Oho 1-1, Tsukuba, Ibaraki, 305-0801, Japan}

\emailAdd{tomohisa.ogawa@kek.jp}

\abstract{
This paper presents studies on application of convolutional neural network (CNN) to GEANT4 optical simulation data generated with a scintillator detector subdivided into 1 cubic cm, which is designed for the long-baseline neutrino experiment. Classification of interaction, regression of momentum, and segmentation of hits are demonstrated for single particle and neutrino-nucleon interaction events with well established CNN architectures by feeding reconstructed 2D projection images. In the study it is shown that the application of CNN to the 1 cm subdivided scintillator detector can provide a factor about 2 better momentum resolution compared to a standard method, as well as a classification capability of about 94\% for the single particle and 70\% for the neutrino-nucleon interaction events. Cross-section analyses with CNN is also shown to be feasible.
}

\keywords{Neutrino detector, Deep neural networks, Pattern recognition, cluster finding, calibration and fitting methods, Particle identification methods}

\proceeding{
}

\begin{document}

\maketitle

\flushbottom

\section{Introduction}

Discovery and precision measurement of CP violation is one of the strong motivations for future long-baseline neutrino experiments. The discovery of the CP violation will be an important step for explication of the matter-antimatter asymmetry in the early universe that is introduced by a theory of leptogenesis~\cite{PASCOLI20071}. Aiming for the discovery of it, the NOva~\cite{osti_935497} experiment is acquiring data, and T2K~\cite{ABE2011106} is upgrading their apparatuses to reduce systematic errors caused by uncertainty of theoretical neutrino-nucleon interaction models.  



In order to reduce the systematic uncertainty, a fine segmented scintillator detector, called  Super-FGD~\cite{abe2020t2k} for T2K and 3DST for the future DUNE experiment~\cite{Abi_2020}, have been proposed and are under construction. These detectors are consist of a large number of 1 cubic cm scintillator cubes, and allow detailed observation of activity around an interaction vertex and final state particles resulting from the neutrino-nucleon interactions, even for short tracks of low energy hadrons, which have been poorly observed in the past, due to poor segmentation of a detector~\cite{AMAUDRUZ20121}. Thus, trough detailed observations of the neutrino-nucleon interactions, it is expected to improve the systematic uncertainty derived from the theoretical model.

In the field of computer vision, a method called a convolutional neural network (CNN)~\cite{NIPS2012_c399862d}, which can extract features of an image, has been used to achieve remarkable performance in image recognition, regression of observable, object classification, etc. 
In the field of experimental high-energy particle physics (HEP) a general analysis method for event reconstruction and classification is to apply specialized human-tuned algorithms to data taken with detectors. In fact, imperfections of the algorithms can cause biases such as rejection or absorption of subtle physical information originally held in the data.
However, if the data can be directly imported as ``images'' based on the CNN method, processes such as feature extraction of event topology, classification of interactions, and estimation of energy or momentum of an incident particle can be done automatically without using such artificial and complicated algorithms. Therefore, it should be possible to make the best use of underlying information in the data.

In this paper, single-particle and neutrino-nucleon events are generated based on the proposed scintillator detector subdivided into 1 cubic cm, and it is shown that the CNN method can be applied to tasks of classification, regression, and segmentation of those interactions and achieve better performance compared to standard analysis method. Section~\ref{sec:Instruments} introduces instruments for the simulation used in the study. Section~\ref{sec:Architectures} explains CNN architectures and its training. Section~\ref{sec:SingleParticle} and~\ref{sec:NeutrinoNucleon} show performance of CNN for single particle and neutrino-nucleon interactions. Finally, discussion and summary are mentioned in section~\ref{sec:DiscussionSummary}.



\section{Instruments for simulation}
\label{sec:Instruments}

Single particle and neutrino-nucleon interactions were generated using the GEANT4 framework~\cite{AGOSTINELLI2003250} that can simulate interactions of the particle inside the detector model. The NEUT package~\cite{HAYATO2002171} was used for generation of seed samples for the neutrino-nucleon interactions, and those samples are fed into the GEANT4 framework. 

\subsection{A detector model}

Super-FGD for the T2K experiment is, for instance, consist of about two million 1 cubic cm scintillator cubes where scintillation light is readout from three orthogonal directions by wavelength shifting fibers, and the light is detected by a Multi-Pixel Photon Counter (MPPC) attached on the fiber edge.

In the simulation, similar detector model is designed. A 1 cubic cm scintillator cube is stacked in ${\rm 100~cm^3}$, and the wavelength shifting fibers are inserted into those cubes along three orthogonal directions. A pseudo-MPPC is mounted on one end of the fiber to acquire a photon, and record the amount of it and its arrival time. Another end of the fiber remains open without no treatment. The simulation is based on optical processes supported by the GEANT4 framework where each material defined by a user requires a table of material properties for transportation of an (optical) photon.    

 

The main material of the scintillator cube is polystyrene where a refractive index is set to 1.50 as typical value of a plastic scintillator and attenuation length to 35~cm~\cite{1462328}. The scintillator cube is composed of two fluorescent material, which are generally PTP of 1.0\% and POPOP of 0.03\%~\cite{1462328, Amaudruz_2012}. Although the scintillation is physically transported by those two material, relative emission spectra of POPOP~\cite{ADAM2007523} only is implemented with a time constant of 1~nsec in the simulation since it is an observable spectrum eventually. 10 photon/KeV and 0.126 mm/MeV are originally set in GEANT4 as typical scintillation-yield of a practical scintillator~\cite{KANDEMIR201830} and the Birk's constant for the polystyrene. The cube is covered with a reflective layer of 90 um~\cite{mth2019_fujita} with composition of ${\rm TiO_2}$, which is formed by chemical etching so as to optically separate each other. Boundary properties between the cube and reflective layer is set to be UNIFIED surface model~\cite{GATEdocumentation} where the reflective index and refractive index are respectively set to be 0.97 and 2.20.  



\begin{figure}[htbp]
        \begin{center}
        \includegraphics[width=80mm]{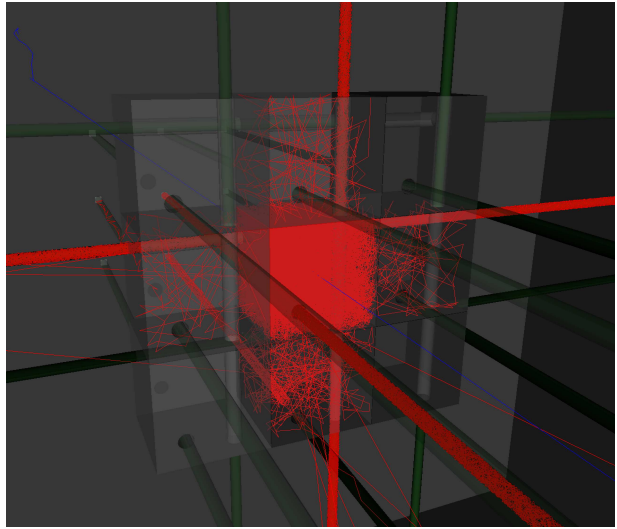}
        \end{center}
\caption{
An example of response of the scintillator cubes stacked in 3$\times$3 when $\mu^-$ (shown in blue) is injected into the center. Propagation of optical photons is shown in red.
}
\label{fig:CUBE}
\end{figure}

The wavelength shifting fiber with a multi-clad is implemented by referring to its specification (Kuraray Y-11(200)~\cite{KurarayWavelength}). A core material of the fiber is pethylene with the refractive index of 1.59, and the attenuation length and relative emission spectrum refers to results~\cite{ADAM2007523}. The inner and outer clad are PMMA and Fluorinated Polyethylene. Each has refractive index of 1.49 and 1.42 respectively. The attenuation length refers to a paper~\cite{ADAM2007523}. MPPC has a photon detection efficiency that depends on a wavelength, and its implementation refers to specification (MPPC S13360 series~\cite{MPPCsS13360}). A magnetic field of 0.2~T is applied to the entire detector in a z direction, assuming SFGD. An example of response in stacked 3$\times$3 scintillator cubes is illustrated in \Figref{fig:CUBE}. 








\begin{figure}[htbp]
        \begin{center}
        \includegraphics[width=150mm]{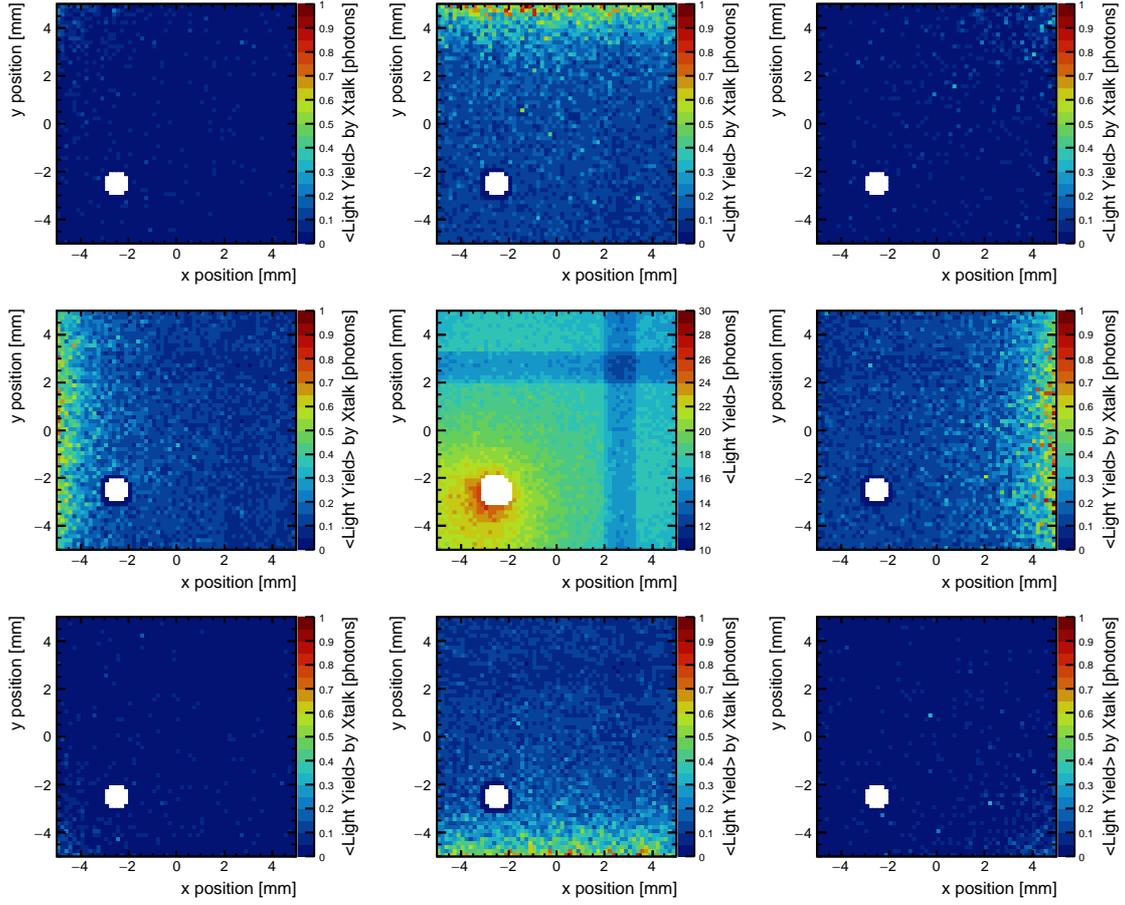}
        \end{center}
\caption{
The figures show a light yield map (center) readout from z direction when a cosmic with a few GeV passed through specific position in the cube along the z direction, and optical crosstalk maps on the cubes neighboring to the center cube.
}
\label{fig:RESPONCECUBE}
\end{figure}


The center plot in \Figref{fig:RESPONCECUBE} is a map showing average light yield readout from z-direction with three inserted fibers when a 1 GeV muon passed through perpendicular (z-direction) in the cube. 
Depending on the number of fibers inserted into the cube, the observed average light yield acquired with the fibers in a particular direction differs, which are respectively 23.2$\pm$0.7, 19.7$\pm$0.6, and 17.5$\pm$0.6 when one, two, or three fibers were inserted, respectively. The light yield gradually decreases with a factor of about 0.15 due to absorption by the closest fiber from a generated point of the photon. This observation is in good agreement with actual measurements using cosmic rays~\cite{mth2019_fujita}.


Although the cube is covered by the reflective layer, the photons generated at certain cube can travel to neighboring cubes due to the hole. This is called cube-to-cube optical crosstalk, which are shown in maps around the center one in \Figref{fig:RESPONCECUBE}. Each point on the map corresponds to a value of the amount of the optical crosstalk readout from the z-direction when the cosmic ray passes through the same position on the center cube. There is almost no optical crosstalk on the diagonal cubes. The optical crosstalk, however, occurs in the adjacent cubes, which is about more than 0.5 photons near the center cube. This is comparable to more than about 2.5\% comparing with the observed light yield in the center cube. In the literature~\cite{Blondel_2020}, 2.94\% is estimated by experiment as a factor for the optical crosstalk. When the injection point of the cosmic ray is farther away, the amount of optical crosstalk decreases to less than 1\%. This value might be slightly smaller compared to the actual experiment referred to the above. This is probably due to the fact that the corners and edges of the cube reduce the thickness of the reflector due to friction between them during the manufacturing and transportation process, however it will be difficult to implement such conditions in the simulation model.


\subsection{Event generation}
 
 

To train CNN architecture, single particle events of electron:$e^-$, gamma:$\gamma$, muon:$\mu^-$, pion:$\pi^-$, and proton:$p$ are generated with the detector model. The particles are isotropically generated from the center of the detector. Momentum of the particles injected into the cube are uniformly distributed in a range of 300 to 700~MeV by changing kinetic energy randomly. This low momentum range can be newly covered by the proposed scintillator detector with the segmentation of 1 cubic cm.  



Physics processe registered in the simulation is FTFP\_BERT~\cite{AGOSTINELLI2003250, G4PHYSICSLISTS} in addition to the optical processes, which is a recommended package by GEANT4 for high energy physics experiments. It can handle not only electromagnetic processes but hadron processes for $\pi$, $p$ and neutron:$n$ in the neutrino-nucleon interactions. 



When acquiring the light yield with MPPC, a hit is defined in which a requirement that the light yield must not be less than three photos is imposed. This is based on the assumption that random dark noise from MPPC should be rejected by this constraint. A 3D-position of a hit-object is defied by taking an intersection of MPPCs with the light yield acquired in the three directions. After collecting all hit-objects, a 3D image is provided, and each coordinate plane provides projected images: xz, yz, and xy for each event. 



The neutrino-nucleon interactions are also generated based on the same detector model, and the CNN architecture is trained. A seed neutrino-nucleon interaction is generated by NEUT where an incident neutrino is assumed to be $\nu_\mu$ or $\nu_e$ and their energies are uniformly distributed from 500 to 1500~MeV (except a few processes, mentioned later), which corresponds to a typical energy range of the T2K experiment.  

The neutrino-nucleon interactions considered in the study are only neutrino-nucleon charged-current (CC) interactions, which leave a charged lepton in the final state. Since neutral-current (NC) interactions are sufficiently separated in past T2K analysises by identifying isolated leptons~\cite{t2kcollaboration2021improved}, the NC interactions are not focused on here. The CC interactions are further divided into the following sub-processes.



 
\begin{enumerate}

\item Quasi-elastic scattering (CCQE) : With an incoming neutrino having insufficient energy to break a nucleon, this process happens via exchange of an intermediate vector boson: $W^+$. The process is a typical two-body system, with a proton recoiling from a scattered lepton. However, re-scattering of the proton inside the nucleus is possible to produce multiple baryons: protons and neutrons in the final state. This final state interactions (FSI) makes it particularly difficult to separate the CCQE process from the following sub-processes.      
           

     
\item Two nucleons producing two holes (CC2p2h) : Two nucleons are involved in this interaction. In the current analysis, it is thought to be contaminating CCQE with a ratio of CC2p2h/CCQE = 10\%-20\%. However, the rate is not known well, and one of the purposes of introducing the proposed detector is to reveal the extent of the contamination with the first sub-process and to reduce the systematics due to uncertainty of both interactions. The FSI occurs as well, making the separation further challenging.

    
      
\item Single $\pi$ production via delta resonance (CC1$\pi$) : This process is dominated when the neutrino energy is about over 1~GeV, and produces a single pion via decay of baryonic resonance. Since energy of the final state lepton cannot be reconstructed well due to decay chain of the pion, this process is background against CCQE. It is important to divide this process sufficiently from the previous two sub-processes to reduce the systematics of the model. Three kinds of production via the interaction with $p$ or $n$ are considered: $l^{-}+p+\pi^+$, $l^{-}+p+\pi^0$, and $l^-+n+\pi^+$ 
       
        

\item Single $\gamma$, $K$, and $\eta$ production via delta resonance (CC0$\pi$) and multi-pion production (CCN$\pi$) : Other hadrons are produced via the baryonic resonance when the neutrino energy gets higher, and multiple pions appear in the final state. These sub-processes are background for the single-pion sub-process and need to be clearly separated.



\end{enumerate}


In preparation of training the networks, each image is cropped to a 100$\times$100 ${\rm cm^2}$ pixel (cube) size. This procedure is indispensable to avoid constraints of GPU memory, time consumption for the training, and over-training. The center of the image for the cropping is decided by taking the center of the light yield (energy deposit) $\vec{X}_{\rm c.o.g}$ as follows,
\begin{eqnarray}
  \vec{X}_{\rm c.o.g} = 
        \frac{ {\displaystyle\sum\nolimits_{i=0}} ~ \vec{x}_i \cdot {\rm{LY}}_{\vec{x}_i} }
                { {\displaystyle\sum\nolimits_{i=0}} ~                       {\rm{LY}}_{\vec{x}_i} }   
\end{eqnarray}
where $\vec{x}_i$ is the 3D-position of the $i$-th hit-object, and $\rm{LY}$ is the observed light yield in the corresponding readout plane. After cropping the images, the events are divided into a training dataset of 80\% and validation dataset of 20\% of the total data. The validation dataset, which does not share any events with the training dataset, is used to monitor the hyper-parameters and determine whether the training results are good or bad. As the data, 100,000 events are prepared for each single particle and sub-processes of the neutrino-nucleon interaction.

\section{CNN architectures and training}
\label{sec:Architectures}

CNN, which performs particularly well in image classification and pattern recognition, has been shown to be applicable to a wide range of tasks in a variety of fields, and their  applications have extended to the field of HEP: online triggering~\cite{Iiyama_2021}, hadron jet and electromagnetic shower clustering~\cite{de_Oliveira_2016}, event classification and background suppression via its topology~\cite{Renner_2017}. In the field of long-baseline neutrino experiment, NOvA and MicroBooNE collaborations have demonstrated the  classification performance for events of the neutrino-nucleon interactions~\cite{Aurisano_2016, Acciarri_2017}.

Since the detector can provide 3D information of the hit-object, the best approach is to directly exploit the hit-objects, which contain all underlying information to be exploited, and 3D CNN has been proposed in several literatures for this purpose~\cite{Ai_2018}. However, the direct usage of the 3D information requires a huge amount of GPU memory, however, the distribution of hit-objects obtained by the neutrino-nucleon interaction is in fact very sparse. Therefore, 3D CNN is not effective method here, and the study shown in this paper is done with standard 2D CNN.


\subsection{Architectures}

In the study, two kinds of CNN architectures are used, which are well established and have been applied to several tasks in HEP. 


One is GoogLeNet~\cite{DBLP:journals/corr/SzegedyLJSRAEVR14}, which is characterized by a small network called an Inception module, which convolutes images in parallel with different sizes and its outputs are concatenated together. This allows to go deeper in the number of layers without increasing the number of parameters before the image becomes too small to convolute. Thus the architecture can get diversity.


Another is U-Net~\cite{DBLP:journals/corr/RonnebergerFB15}, which was originally developed for segmentation of a cell in the biomedical filed. U-Net extracts features by convoluting the image, similar to CNN, but instead, position information gets ambiguous in CNN. However, U-Net composes a deconvolution process and concatenates the convoluted feature maps formed in the same dimensions after the deconvolution. This process makes it possible to extract the features of the image while retaining the position information.




\subsection{Training}

%
%


The CNN architectures are implemented and trained using Tensorflow (2.4.1)~\cite{Tens} and Keras (2.4.0)~\cite{Keras} on python. For the training the projected images (xy, yz, xz) of each event is input into the network in three channels which is analogous to RGB of a color image. Here, there are two means when inputing the image: xy, yz, and xz together or separately. Some studies follow the former~\cite{Renner_2017, Acciarri_2017} and others follow the latter~\cite{Aurisano_2016}. When testing $\mu^-\mhyph\pi^-$ classification with both means based on GoogLeNet, both gives similar performance: together one is 94.43$\pm$0.11\% and separately is 94.39$\pm$0.11\%. Thus, the former is selected in the following studies since the latter requires three times parameters as much as the former.     




The training of the networks was done in an environment provided by Google Colaboratory~\cite{GoogleColab} where GPUs are available for free for deep learning related studies. The computation time available for free in this environment is limited, nevertheless, it is sufficient to demonstrate some studies and estimate its performances. With an image resolution of 100$\times$100 for three projections, and a batch size of 64, the training speed reached about 280 msec/batch with U-Net having 10~M trainable parameters on a single NVIDIA Tesla P100 16GB PCI-e (and about 430 msec/batch on a single NVIDIA Tesla T4 with 16GB memory).   

Total size of the data for the tasks was over 100 GB. Due to the limited disk space in the environment, the data is divided into several sub-datasets and the training was done by replacing those datasets. To accommodate the replacement of the datasets, a learning rate was gradually reduced by a factor of 0.8 if there were no updates in several epochs. This procedure can make sense since each dataset has the same true minimum solution. Data augmentation technique is also applied to compensate for the amount of the training data.

As the loss function for the classification task, the probability of each category is calculated using the categorical cross entropy along with the softmax function~\cite{NIPS2012_c399862d} defined as follows. A batch size averaged relative squared error, also defined as follows, along with the linear function is used for the regression task. The sum of the loss functions is used for the classification and the regression task.
\begin{eqnarray}
L_{\rm class} &=& - 1/m \cdot  \sum\nolimits_{i=0}^{m} \sum\nolimits_{j=0}^{\rm class} 
				~   t^i_{j} \log ( {s}^i_{j} )
				\\ [1mm]
L_{\rm regr} &=& 1/m \cdot \sum\nolimits_{i=0}^{m} 
				~ \Big[ (P_{\rm pred}^i - P_{\rm true}^i ) / P_{\rm true}^i \Big]^2 
				\\ [2mm]
L_{\rm total} &=& \alpha L_{\rm class} + \beta L_{\rm regr}  \label{Ltotal}
\end{eqnarray}
where $m$, $i$, $j$, $t$ and $s$ are, respectively, the batch size, an $i$-th sample, a $j$-th category, the ground truth and a CNN score for each class $j$ with the softmax function. $P$ shows the momentum of the incident particle for the prediction and MC truth, and $\alpha$ and $\beta$ stand for a weight factor for balancing each loss in the training, and simple values of 1 and $10^{-4}$ are respectively set in the studies.





\section{Single particle interaction}
\label{sec:SingleParticle}

\begin{figure}[H] 
       \begin{center}
       \includegraphics[width=152mm]{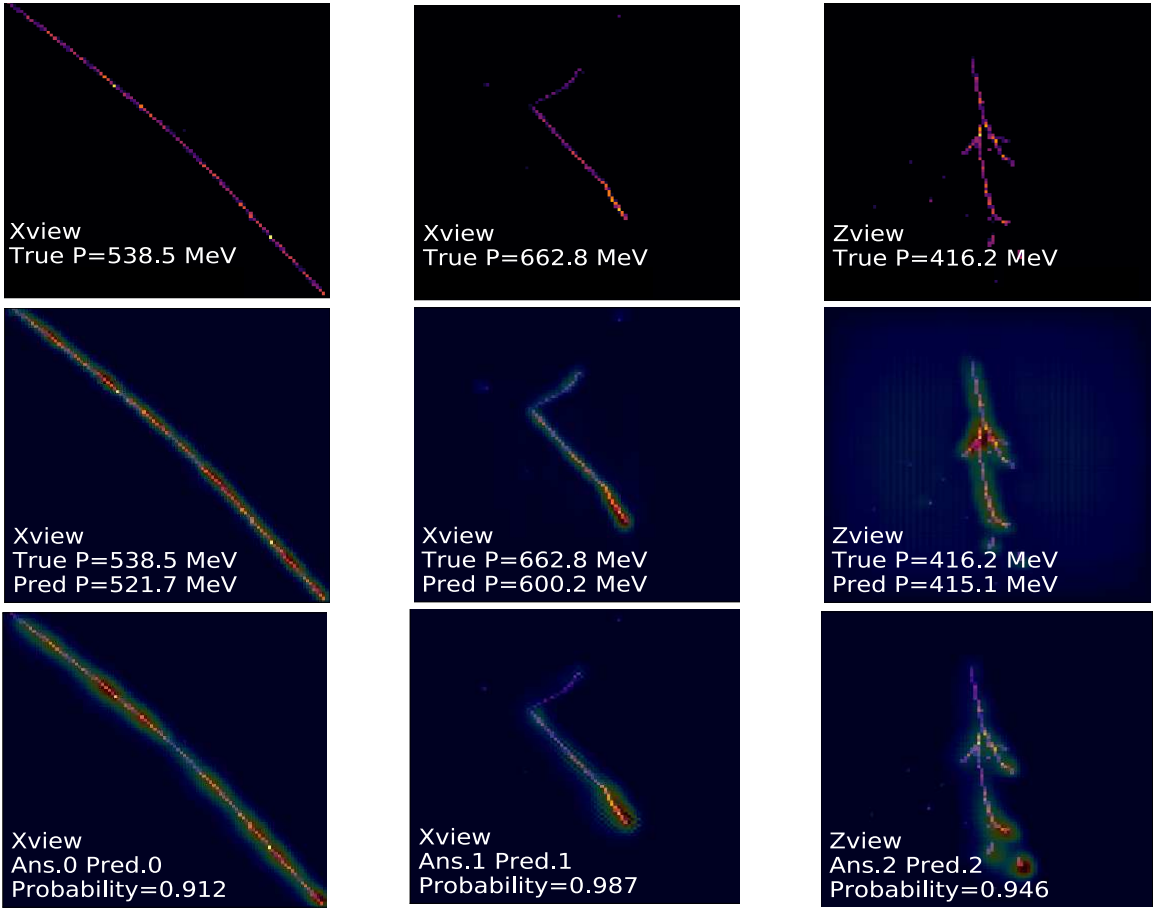}
       \end{center}
\caption{Examples of event displays for $\mu^-$ (left), $\pi^-$ (middle), and $e^-$ (right) interaction. Top three is a hit map projected onto a particular readout plane for each particle, where the color indicates the amount of the light yield. In the middle and bottom, the hit maps are superimposed with a heat map which is colored by the algorithm of Grad-CAM. The network is trained for the regression task for the middle ones and classification task for the bottom ones. True momenta of the incident particles and inference of predicted momenta and classification with probability are shown in the windows. Indices of 0, 1, and 2 in the bottom windows respectively stand for $\mu^-$, $\pi^-$, and $e^-$. 
}
\label{fig:SingleEvent}
\end{figure}

The first row in \Figref{fig:SingleEvent} show examples of hit distributions for each particle type: $\mu^-$, $\pi^-$ and $e^-$ with specific momenta. These distributions are projected onto a certain two-dimensional plane. To train the network, over two hundred epochs was processed to converge the losses. As an inference by the trained network, the regression of the momentum is shown in the second line. The hit distribution in the second line is superimposed with a heat map, which is colored by an algorithm of Gradient-weighted Class Activation Mapping (Grad-CAM)~\cite{wang2020scorecam}, which visualizes the pixels that are important for the network to make predictions. It can be seen that the network tries to focus not only on the pixels with large light yield (energy deposit), but on the overall topology of the interaction.

\subsection{Regression of momentum}



In order to improve the systematic uncertainties arising from the theoretical neutrino-nucleon interaction models, the models need to be minutely understood. Thus, incoming neutrino energy needs to be reconstructed accurately. For example, the momentum of $p$  and the energies of $e^-$ and $\gamma$ from $\mu$ and $\pi$ decays need to be precisely estimated. In the following, $p$ and $e^-$ are focused on. The evaluation of these particles can be a benchmark of the performance of CNN because energy deposit of $p$ is assumed to behave nonlinearly due to the Birk's suppression, and $e^-$ is generally evaluated by photon counting since it is spread by electromagnetic shower. To compare the performance of inference, a standard momentum reconstruction method based on the photon counting was implemented: the number of photons are converted to the number of MIPs, and it is multiplied by the minimum energy deposit, which is 2.2 MeV for the implemented scintillator cube.

\begin{figure}[htbp]
    \begin{minipage}{0.5\hsize}
        \begin{center}
        \includegraphics[width=76mm]{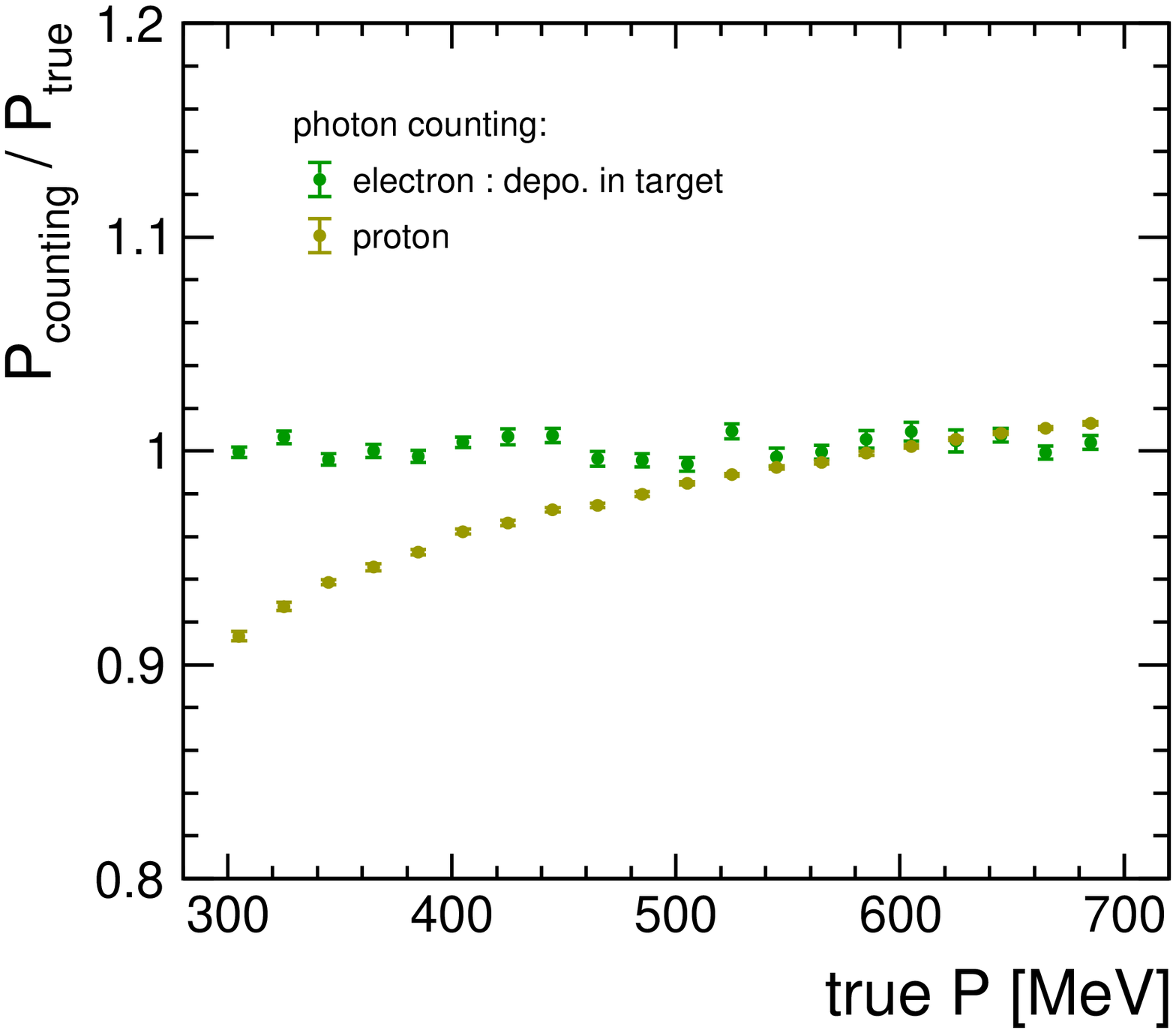}
        \end{center}
    \end{minipage}
    \begin{minipage}{0.5\hsize}
        \begin{center}
        \includegraphics[width=76mm]{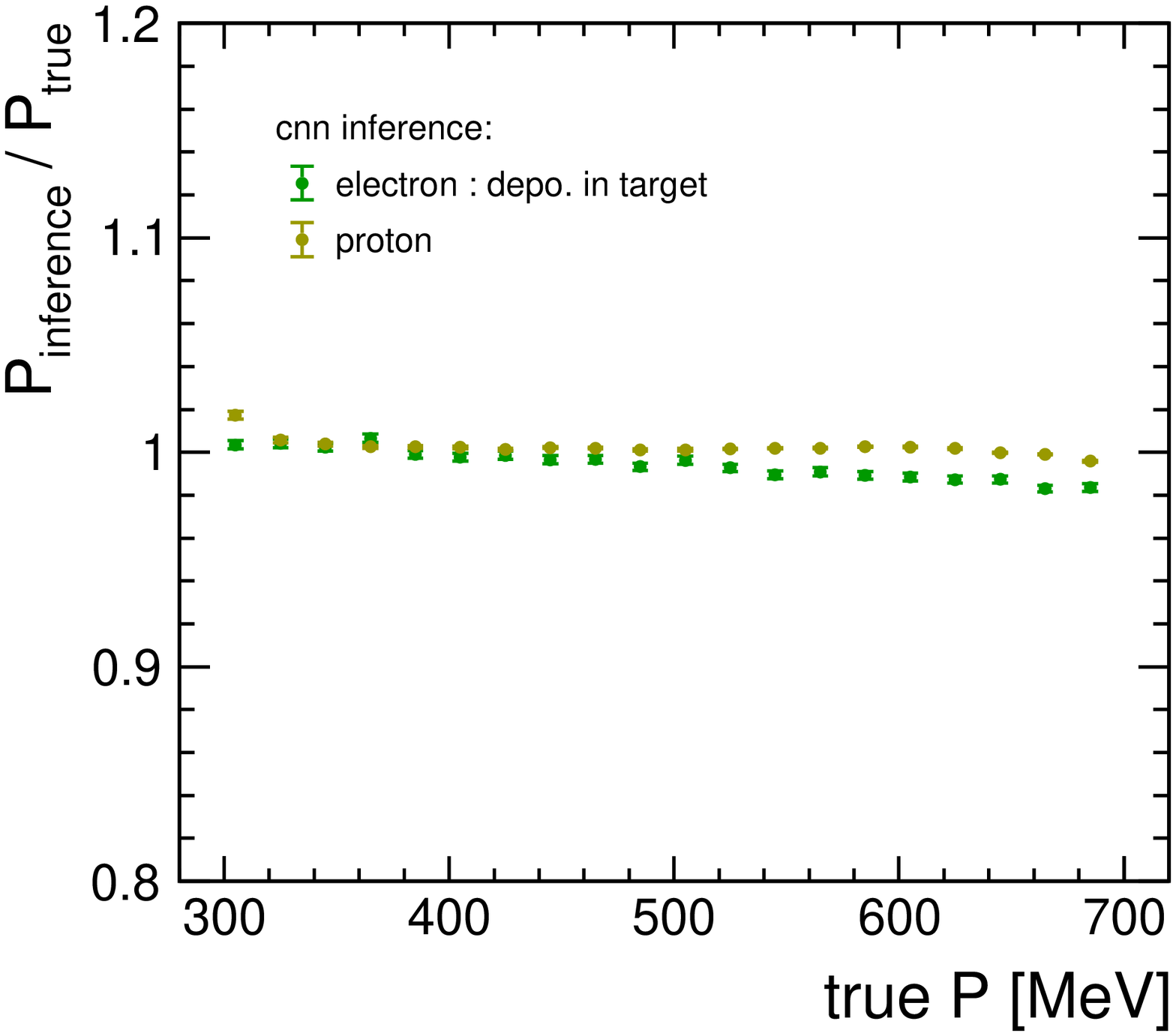}
        \end{center}
    \end{minipage}
\caption{
The ratio of the reconstructed momenta $P_{\rm counting}$ based on the photon counting (left) and inferred momenta $P_{\rm inference}$ with CNN (right) to the true momenta $P_{\rm ture}$ of $e^-$ and $p$. This ratio is quantified in each 20 MeV bin.
}
\label{fig:4_1_1}
\end{figure}

Plots in \Figref{fig:4_1_1} show a ratio of reconstructed momenta based on the photon counting to the true momenta $P_{\rm counting}/P_{\rm ture}$ of $e^-$ and $p$, and predicted momenta to the true momenta $P_{\rm inference}/P_{\rm ture}$ of $e^-$ and $p$ It must be noted that since radiation length of the scintillator is over 40~cm, a fraction of the energy of $e^-$ can easily escape from the detector. Thus, the deposited energy, not the incident energy, is considered as $P_{\rm ture}$ for $e^-$.

In a plot with the photon counting, a nonlinear effect reaches 10\% can be seen at low momentum of $p$ due to large energy deposit and its Birk's suppression while the response of $e^-$ is stably 1 over the given momentum range. In contrast, in a plot with the CNN inference, CNN corrects for the nonlinear effect by looking at a large amount of samples and learning its statistical behavior of $p$. It can be seen that the ratio for $e^-$ is slightly off from 1 at high momentum. This is perhaps due to the combined heterogeneity of the detector model, such as shape of the detector itself with respect to the initial direction of the incident particle, attenuation of the light yield, or threshold effect.

\begin{figure}[H] 
    \begin{minipage}{0.5\hsize}
        \begin{center}
        \includegraphics[width=76mm]{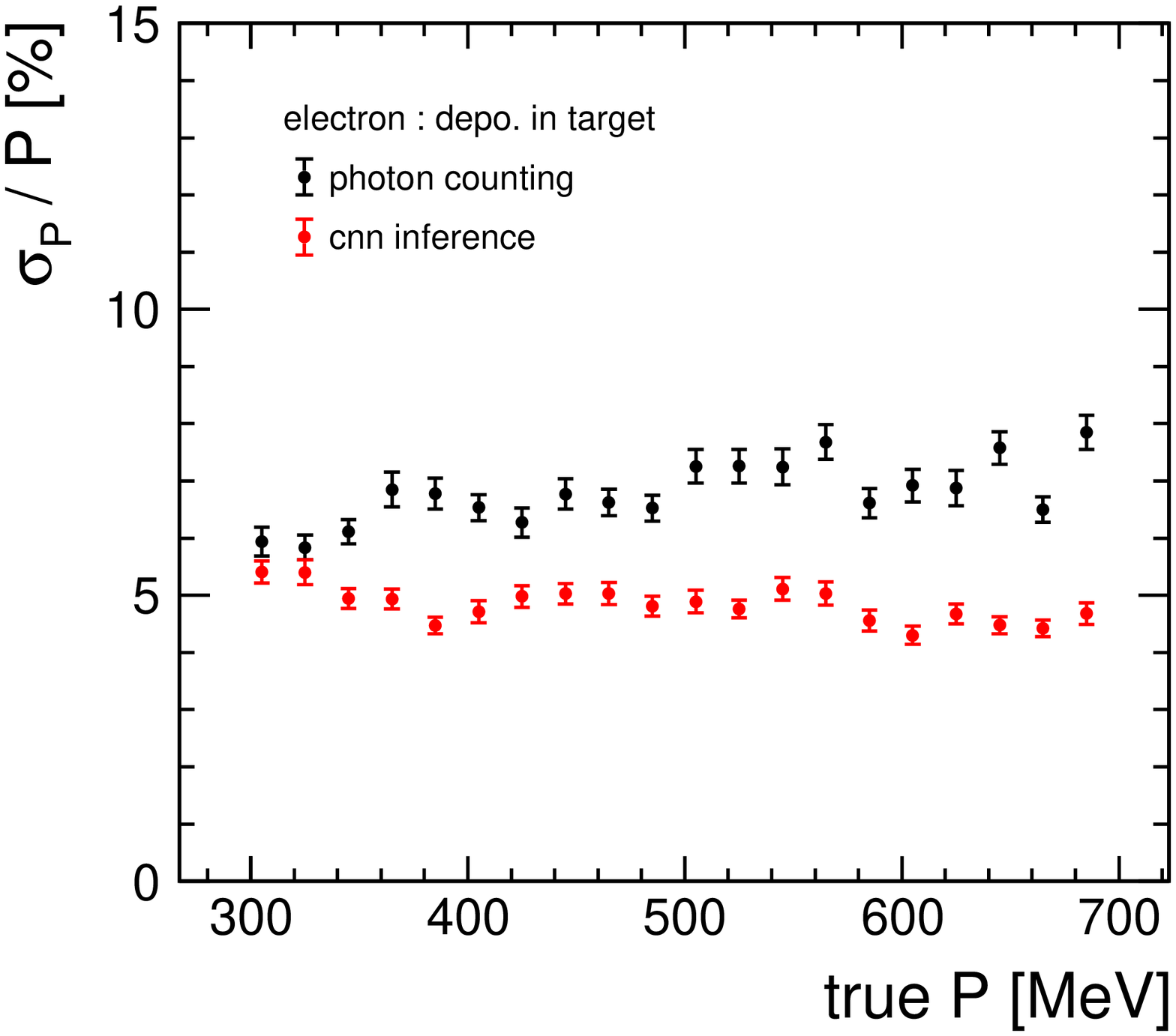}
        \end{center}
    \end{minipage}
    \begin{minipage}{0.5\hsize}
        \begin{center}
        \includegraphics[width=76mm]{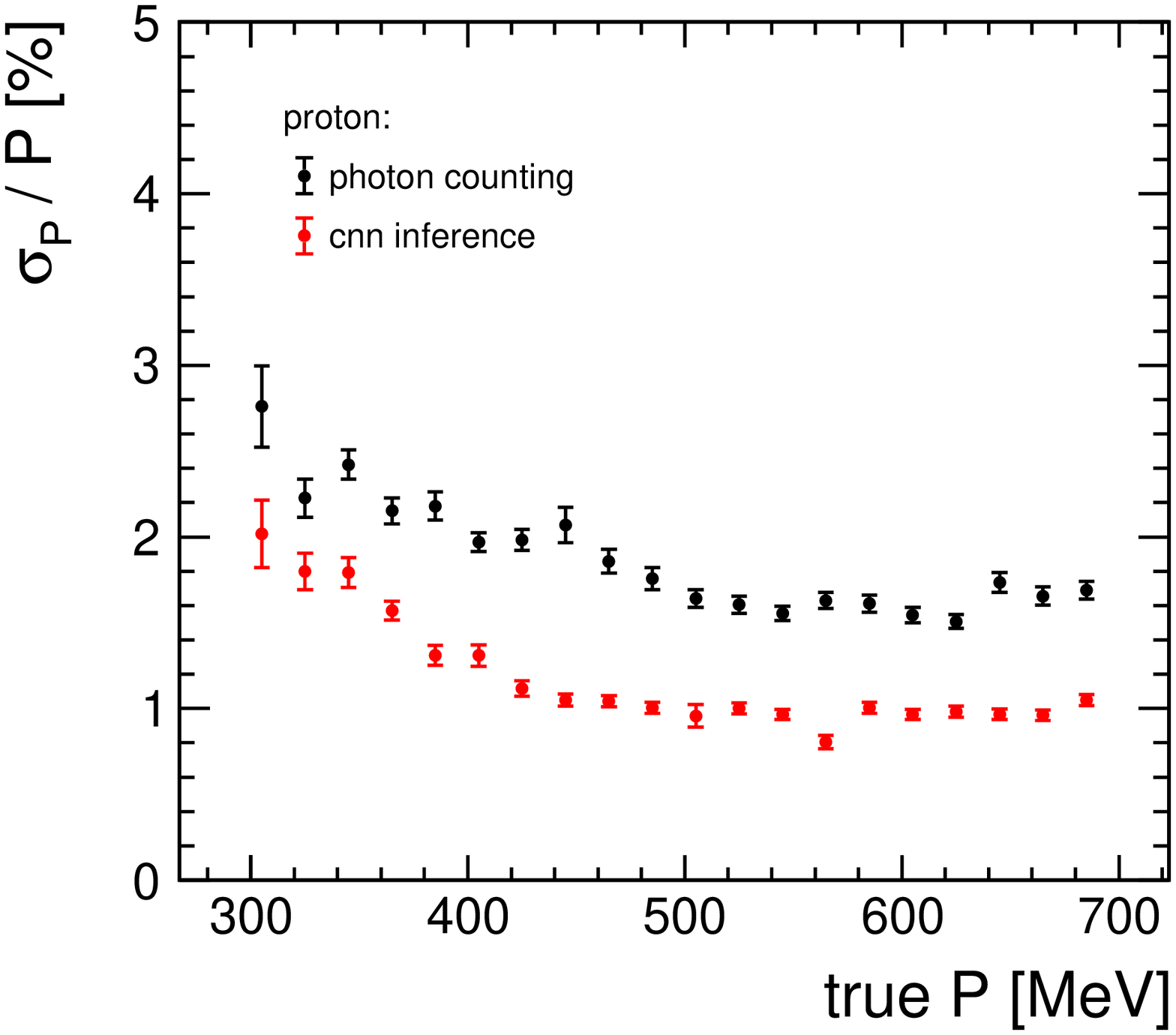}
        \end{center}
    \end{minipage}
\caption{
Momentum resolution of $e^-$ (left) and $p$ (right) by the photon counting and CNN regression. The improvement by CNN is remarkable respectively.
}
\label{fig:4_1_2}
\end{figure}


  
Plots in \Figref{fig:4_1_2} show momentum resolution for $e^-$ and $p$ with both methods. The resolution based of the photon counting for $e^-$ is about 7\%, which could be a reasonable value assuming that about two charged particles via pair creation passed through one scintillator cube: $\sigma_E/E  \propto 1/\sqrt{N} = 1/ \sqrt{17.5\times3\times2} \sim 9.7\%$ (Since the radiation length is large, a large number of charged particles will not pass through one cube). In contrast, the resolution with CNN is about 4\%, which is better with factor of 2. The similar improvement can be seen for the momentum resolution of $p$.

\subsection{Classification of single particles}

The classification was performed by training the architectures using mixed dataset: $\mu^-\mhyph\pi^-$, $e^-\mhyph\gamma$, and five different particles. A summary of the inference is shown in \Tabref{tab:SingleClass}. In the $\mu^-\mhyph\pi^-$ classification both networks give similar inference. In contrast, U-Net achieves higher accuracy in the $e^-\mhyph\gamma$ classification than that by GoogLeNet. This implies that the position information is more important in the $e^-$-$\gamma$ classification. 

\begin{table}[H]
  \begin{center}
    \caption{
Classification accuracy of each mixed dataset is shown as an inference of the trained network. Two different CNN architectures are used and over 300 epochs was processed to get the converged results. LY and TM stand for the light yield and time.
}
\vspace{3mm}
    \begin{tabular}{l|l|c} 
      Classification, Network & \#of layers, parameters, input & Avg. accuracy (\%) \\ \hline \hline
      $\mu^-\mhyph\pi^-$, GoogLeNet & ~~~39,~ 1.7M,~ LY        & 94.43$\pm$0.11  \\
      $\mu^-\mhyph\pi^-$, GoogLeNet & ~~~39,~ 1.7M,~  LY+TM & 94.21$\pm$0.11  \\       
      $\mu^-\mhyph\pi^-$, U-Net         & ~~~41,~ 9.8M,~  LY   & 94.12$\pm$0.11  \\ \hline
      $e^-\mhyph\gamma$, GoogLeNet & ~~~39,~ 1.7M,~  LY   & 94.49$\pm$0.12  \\
      $e^-\mhyph\gamma$, U-Net         & ~~~41,~ 9.8M,~  LY   & 95.04$\pm$0.11  \\ \hline
      5 particles, GoogLeNet                     & ~~~39,~ 1.8M,~  LY  & 93.56$\pm$0.12  \\
      5 particles, U-Net                             & ~~~41,~ 10.M,~  LY  & 94.41$\pm$0.11  \\ \hline
    \end{tabular}
    \label{tab:SingleClass}
  \end{center}
\end{table}




As a test, information of arrival time on the MPPC is also introduced here. For the introduction, each of the two information is standardized and input into the architecture. However, the result does not improve, or even worsened. This is probably because inputting less useful information makes it easier for learning to fall into places like the local minimum.

The third row in \Figref{fig:SingleEvent} shows the heat map when the network is trained to adapt to the classification task. Compared to the regression task, it can be seen that the network is trying to look at a wider area, focusing on the areas with the highest energy deposit. In the case of $e^-$, one or two single hits with low light yield seem to be ignored, while the network seems to take into account the entire topology of the interaction, including out-lands and small fragments.



\begin{figure}[H] 
    \begin{minipage}{0.5\hsize}
        \begin{center}
        \includegraphics[width=76mm]{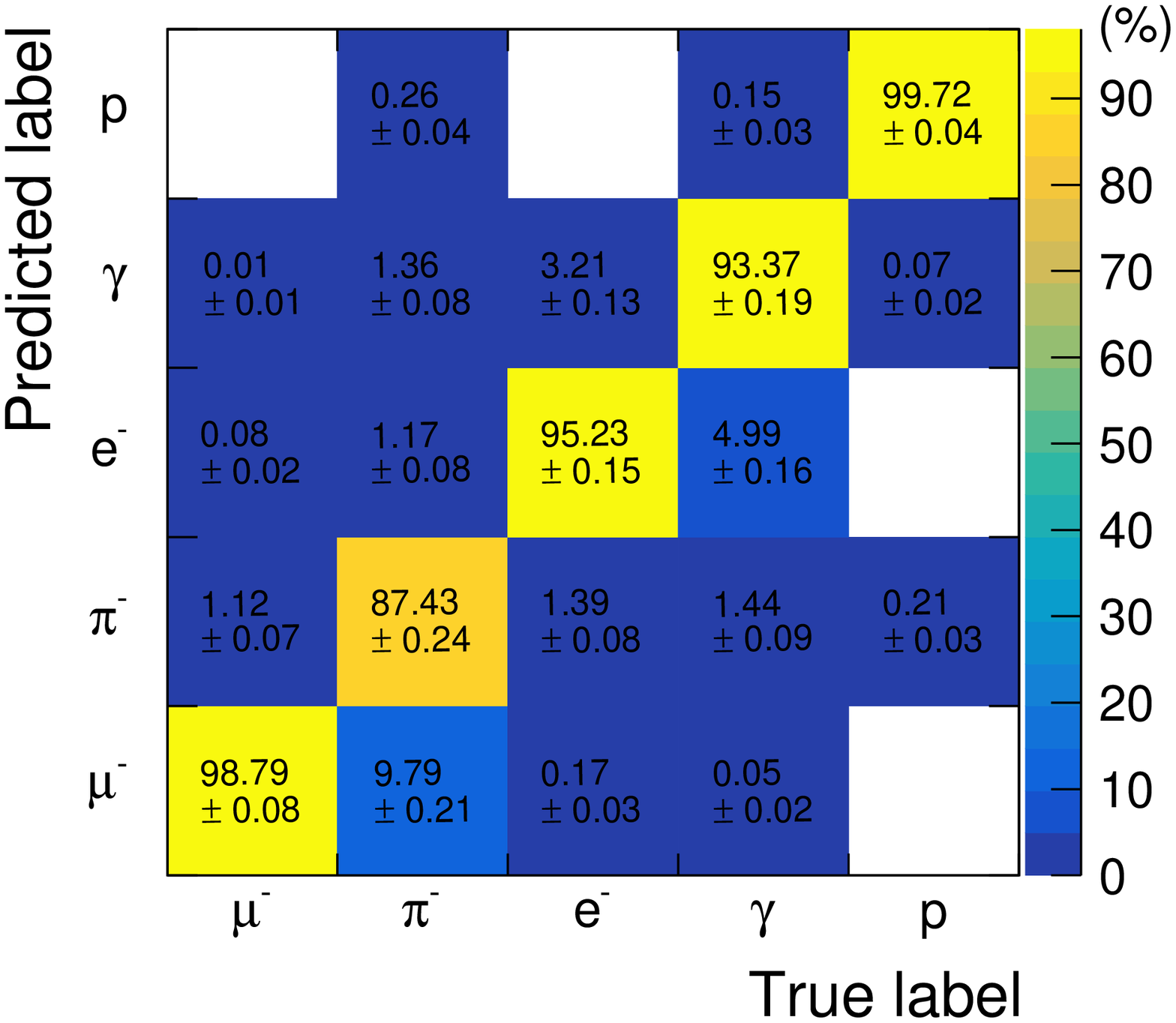}
        \end{center}
    \end{minipage}
    \begin{minipage}{0.5\hsize}
        \begin{center}
        \includegraphics[width=76mm]{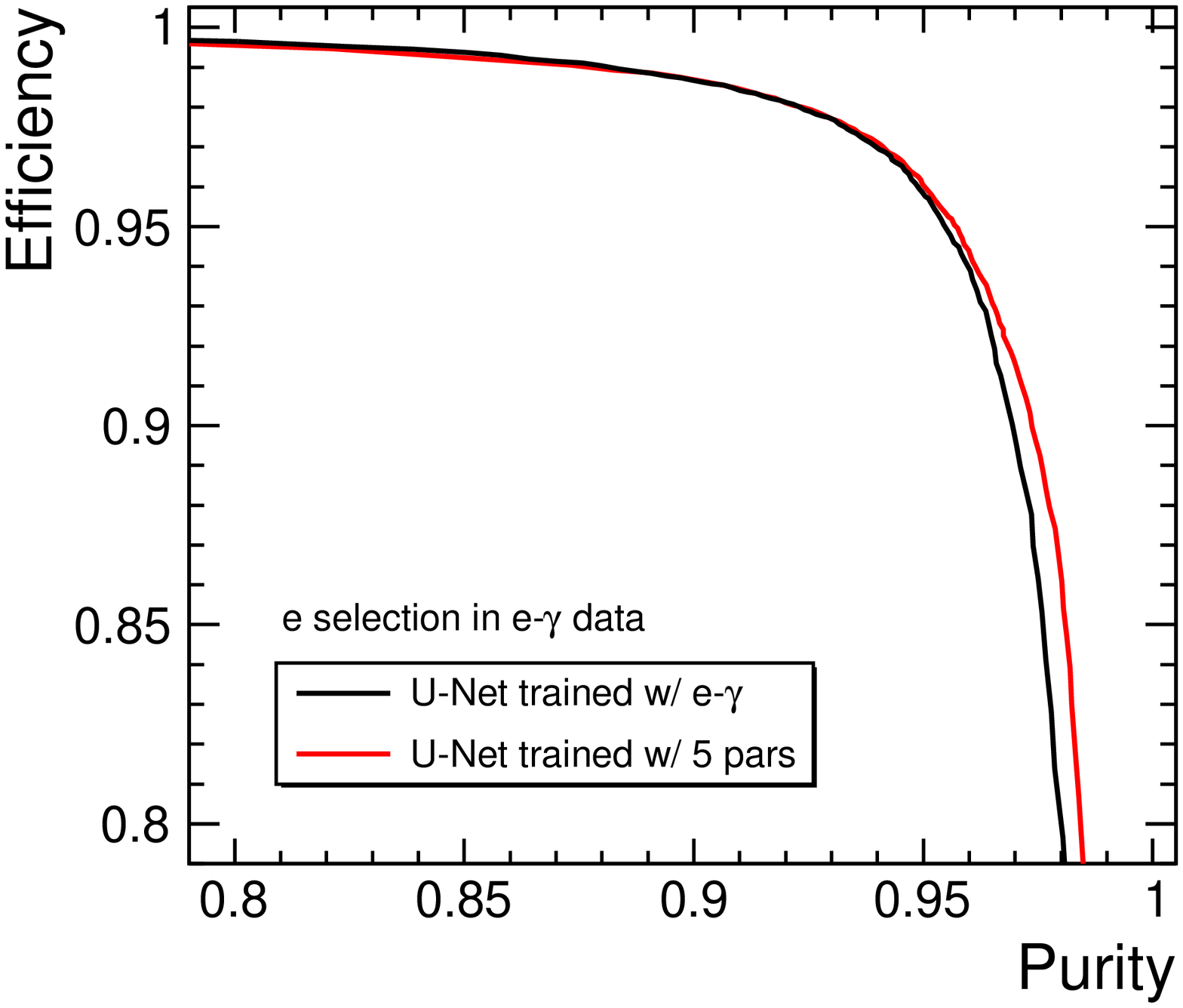}
        \end{center}
    \end{minipage}
\caption{
(Left) A confusion matrix for five different particles when applying U-Net trained by the dataset containing five different particles. (Right) Plot showing the efficiency versus purity curves of $e^-$ when comparing two networks trained with the dataset of $e^-$-$\gamma$ and five different particles.  
}
\label{fig:ConfMat}
\end{figure}

A left plot in \Figref{fig:ConfMat} shows a confusion matrix after performing the classification of five different particles with U-Net. $p$ is clearly separated among the classes, and less confusion happens between $e^-$ and $\gamma$ although the classification accuracies can reach close to 95\%. Some $e^-$ and $\gamma$ are classified as $\pi^-$ by mimicking hadronic shower. The matrix in \Figref{fig:ConfMat} also indicates that relatively large confusion happens between $\mu^-$ and $\pi^-$. $\pi^-$ sometimes looks almost like $\mu^-$, which can happen when $\pi^-$ does not interact with a nucleon and leave kinks. In contrast, $\mu^-$ is not classified as $\pi^-$ becuse it does not often decay while emitting a Michel electron in the given momentum range, leaving no kinks. A right plot in \Figref{fig:ConfMat} gives selection efficiency versus purity (or background suppression) curve, which are defined as follows,


\begin{eqnarray}
{\rm efficiency}(x) &=& \frac{  {\rm true } \cap {\rm select } (p ; p>x) }{   {\rm true}  }, \\ [3mm]
{\rm purity}(x) &=& \frac{  {\rm true } \cap {\rm select } (p ; p>x) }{   {\rm select } (p ; p>x)   } .
\end{eqnarray}
$x$ is certain threshold value to decide the efficiency and purity at a particular point, and $p$ is the probability given by the network for the target class. 

In the right plot, two kinds of U-Net are tested, which are trained with the dataset of $e^-\mhyph\gamma$ and five different particles, respectively. It turns out that the network trained with five different particles gives better performance than the one trained with $e^-\mhyph\gamma$ dataset only. This would be because of the fact that the network exposed to other different particles can acquire a greater diversity of particle interactions and is able to extract more generalized features.



\begin{figure}[H] 
     \begin{minipage}{0.5\hsize}
        \begin{center}
        \includegraphics[width=76mm]{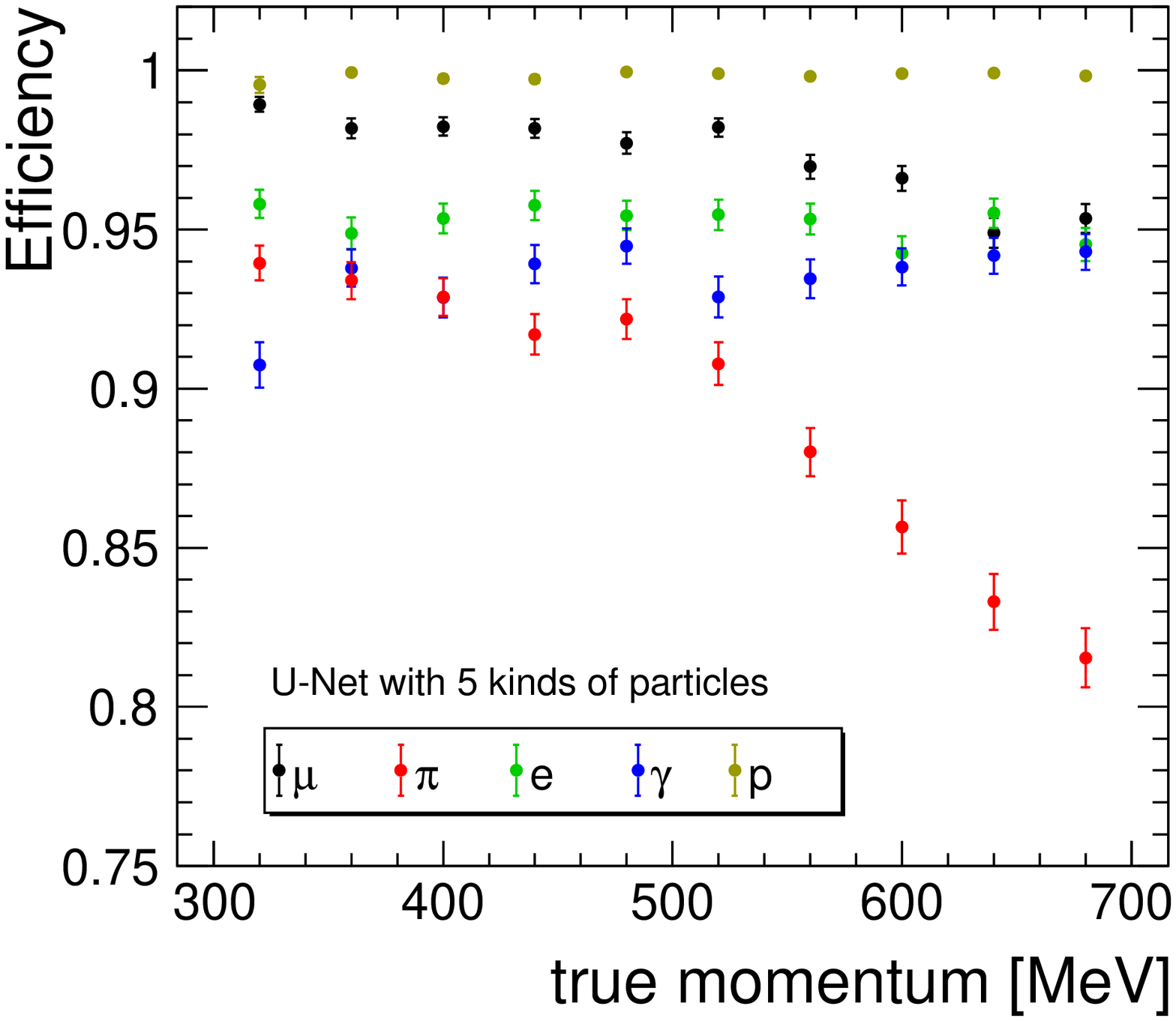}
        \end{center}
    \end{minipage}
    \begin{minipage}{0.5\hsize}
        \begin{center}
        \includegraphics[width=76mm]{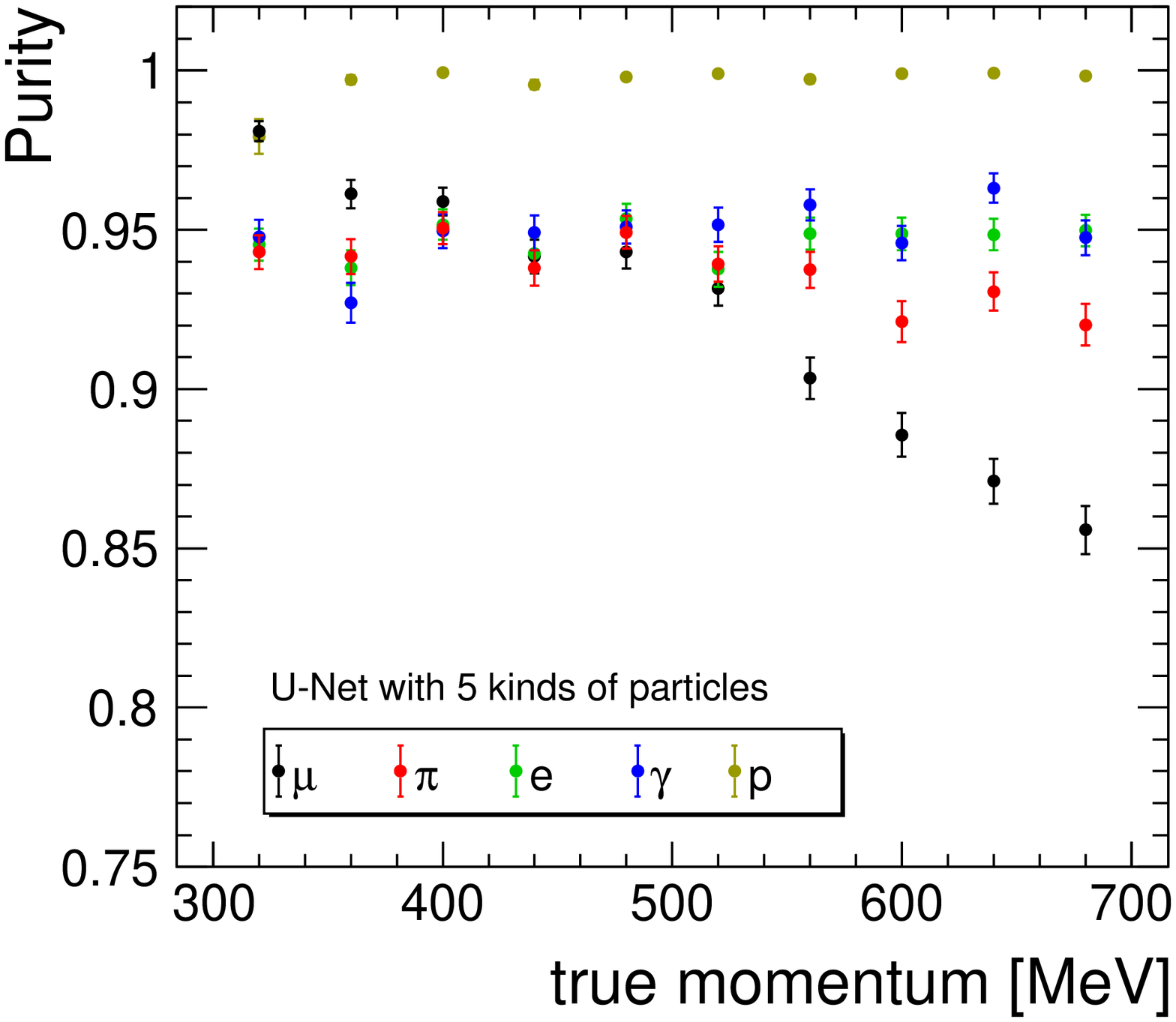}
        \end{center}
    \end{minipage}
\caption{
Plots showing the efficiency and purity as a function of the true momentum of the incident particles The classification is performed by U-Net trained with the dataset of five different particles. The x axis is quantified in each 40 MeV bin. 
}
\label{fig:4_1_20}
\end{figure}

Plots in \Figref{fig:4_1_20} show the efficiency and purity as a function of the true momentum of the incident particles. The momentum is divided into each 40~MeV bin, and the efficiency and purity are optimized with the threshold $x$ in terms of a point where distance between (purity, efficiency) and (1,1) takes a minimum value (the outer most part of the curve). For $e^-$, $\gamma$ and $p$, the plots show that the performance are stable over the given momentum range. However, the purity for the $\mu^-$ classification degrades as the momentum gets large. Since a high energy $\pi^-$ can move longer distance before decay and be less likely to interact with the nucleon besides very similar dE/dx, $\pi^-$ looks closer to $\mu^-$ and is harder to distinguish each other. Therefore, the network classifies it as $\mu^-$ and the efficiency for the $\pi^-$ classification gets worse, similarly the purity for the $\mu^-$ worse. If the information of the external detectors is included in the evaluation, it is expected that the performance for $\mu^-$ and $\pi^-$ can be recovered to more than 90\%.

\section{Neutrino-nucleon interaction}
\label{sec:NeutrinoNucleon}

\begin{figure}[H] 
        \hspace{-3mm}
        \includegraphics[width=156mm]{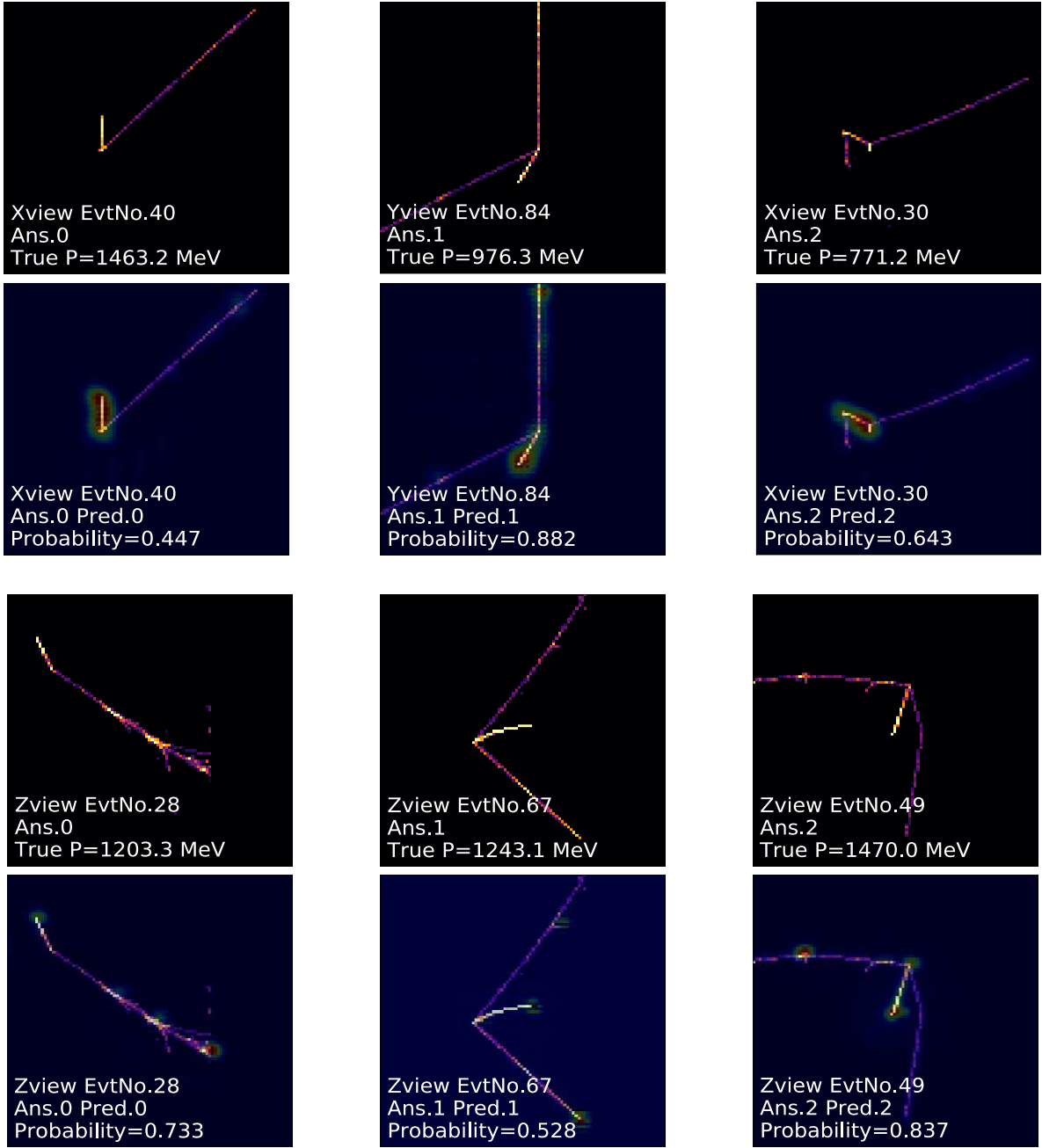}
\caption{Examples of an event display of the neutrino-nucleon interactions: top is $\nu_\mu$ and bottom is $\nu_e$. The process is CCQE, CC2p2h, and CC1$\pi$: $l^{-}+p+\pi^+$ with indices of 0, 1, and 2 from left to right. The heat map is colored by the Grad-CAM algorithm after being trained in the process listed above including CC1$\pi$: $l^{-}+p+\pi^0$ and $l^{-}+n+\pi^+$. The inference of the classification using U-Net is also shown along with its probability. 
}
\label{fig:5_0_1}
\end{figure}

The first row in \Figref{fig:5_0_1} show typical hit distributions of the neutrino-nucleon interactions with $\nu_\mu$, and the third row is with $\nu_e$. The processes are CCQE, CC2p2h, and CC1$\pi$: $l^{-}+p+\pi^+$ from left to right. In the second and fourth rows the heat map by the Grad-CAM algorithm is superimposed on the hit map where U-Net was trained over a few hundred epochs using 5 types of the neutrino-nucleon interactions of, in addition to the listed three interactions above, CC1$\pi$: $l^{-}+p+\pi^0$ and $l^{-}+n+\pi^+$ until the loss converges. Here, it should be noticed that U-Net is used for the following studies since it can provide the better performance than that of GoogleNet. In addition, the training of the classification and regression was perform separately, meaning that the minimization of each term of $L_{\rm total}$ in \Equref{Ltotal} is done separately, to save the training time.

\subsection{Segmentation of hits}

For the purpose of particle reconstruction by clustering detector hits, DB scan or Cellular Automaton have been often used. On the other hand, CNN that has an Encoder-Decoder network structure, such as U-Net, can be also trained by attaching labels to the hits themselves, and can recognize a cluster of the hits that form a characteristic object, meaning the particle, with high accuracy.

\begin{figure}[H]
    \begin{minipage}{0.5\hsize}
        \begin{center}
        \includegraphics[width=76mm]{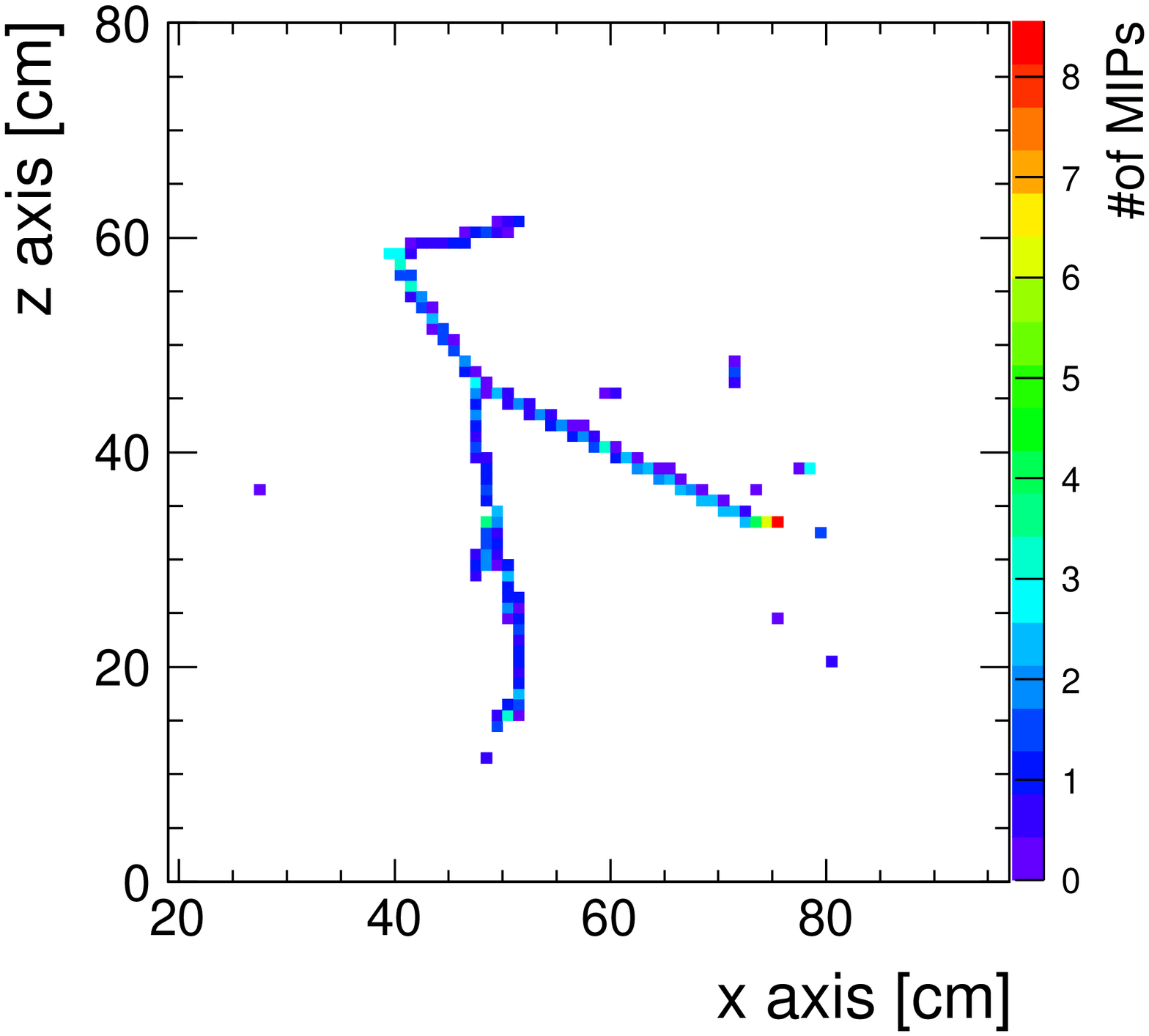}
        \end{center}
    \end{minipage}
    \begin{minipage}{0.5\hsize}
        \begin{center}
        \includegraphics[width=76mm]{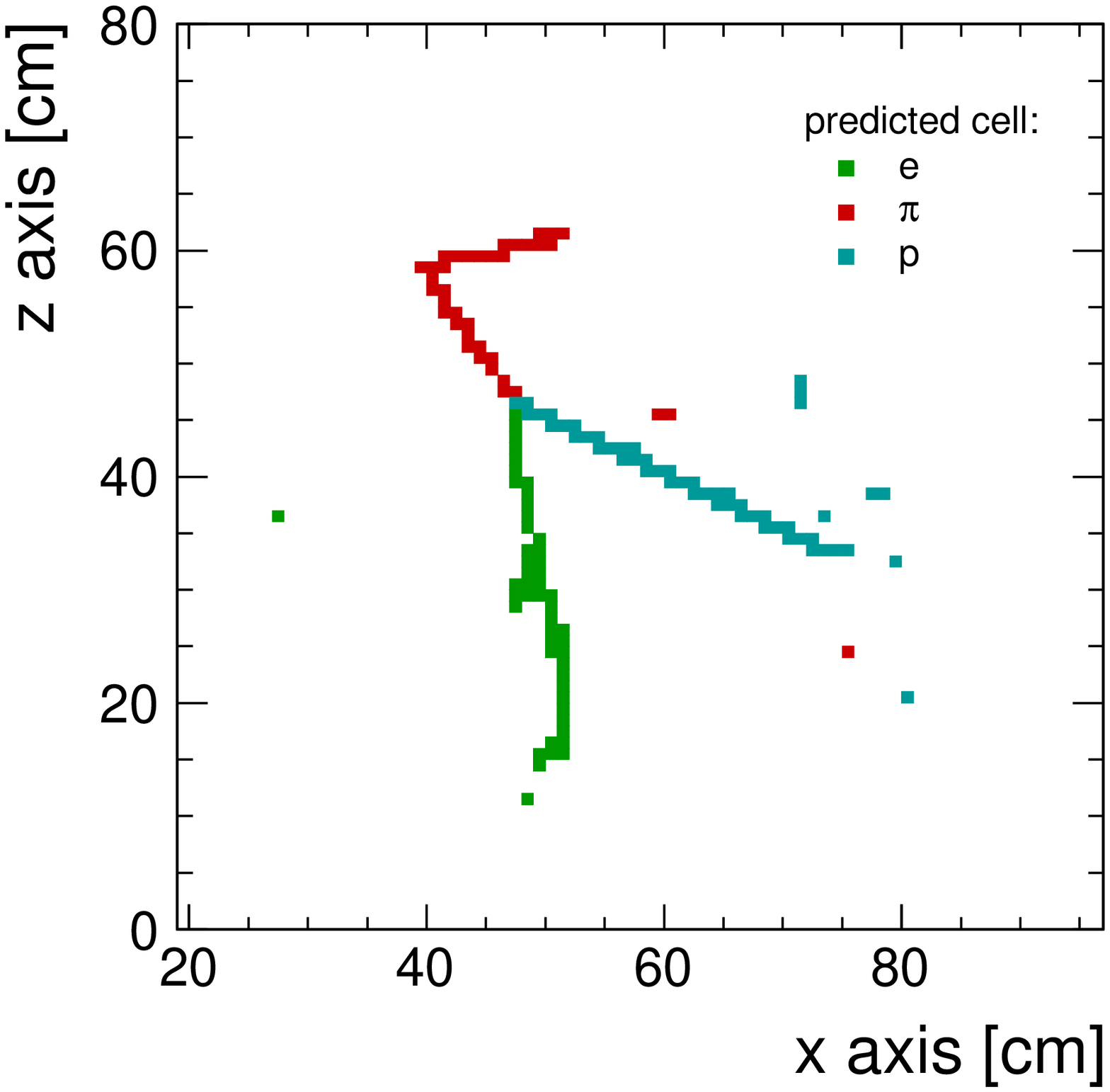}
        \end{center}
    \end{minipage}
\caption{
(Left) A hit display of the CC1$\pi$: $e^-+p+\pi^+$ interactions. (Right) Ground truth labeled on each hit derived from each particle (class). The class that has the largest light yield is labeled if there exist multiple hits by the different class. 
}
\label{fig:5_1_1}
\end{figure}

\begin{figure}[H]
    \begin{minipage}{0.5\hsize}
        \begin{center}
        \includegraphics[width=76mm]{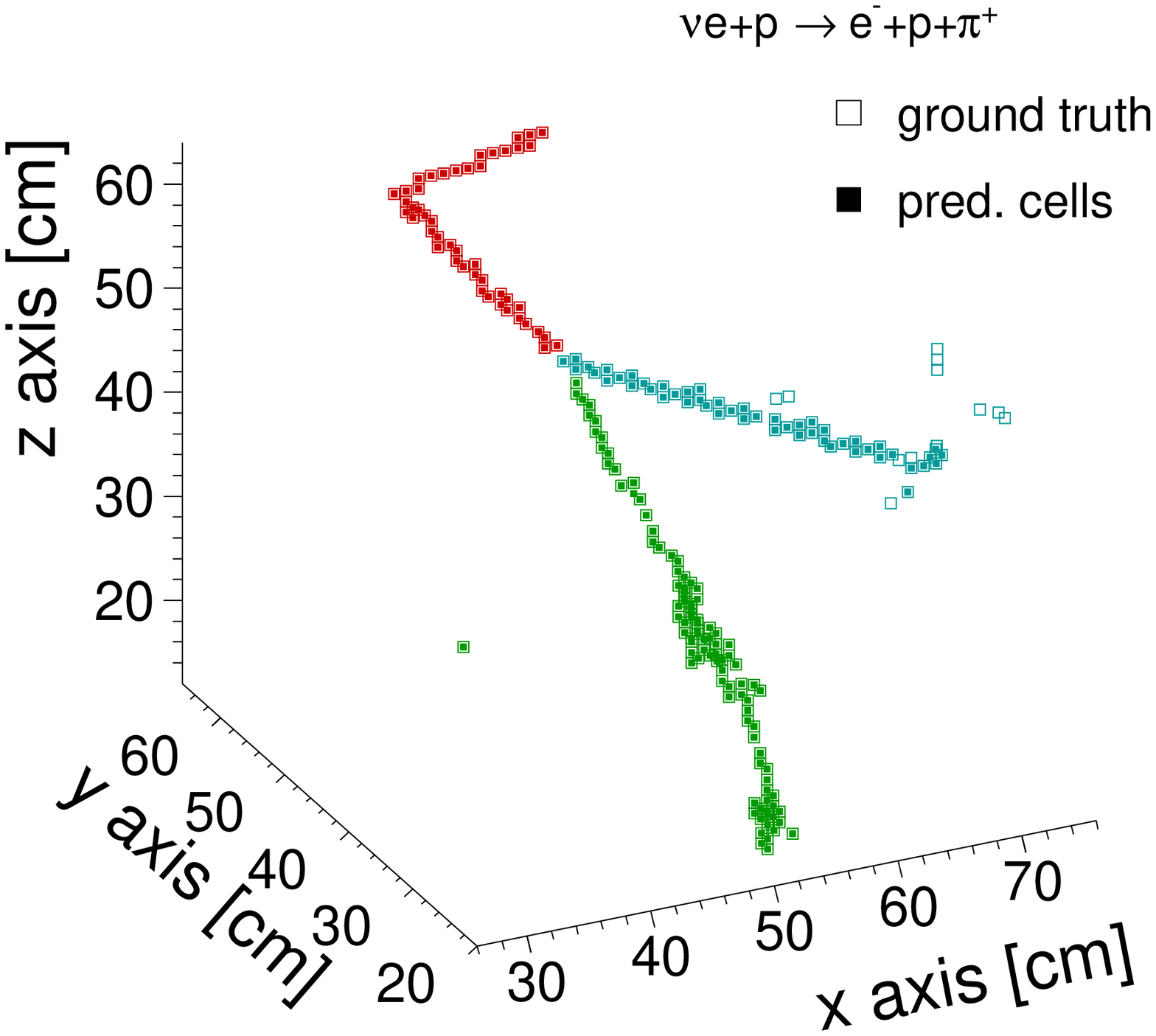}
        \end{center}
    \end{minipage}
    \begin{minipage}{0.5\hsize}
        \begin{center}
        \includegraphics[width=76mm]{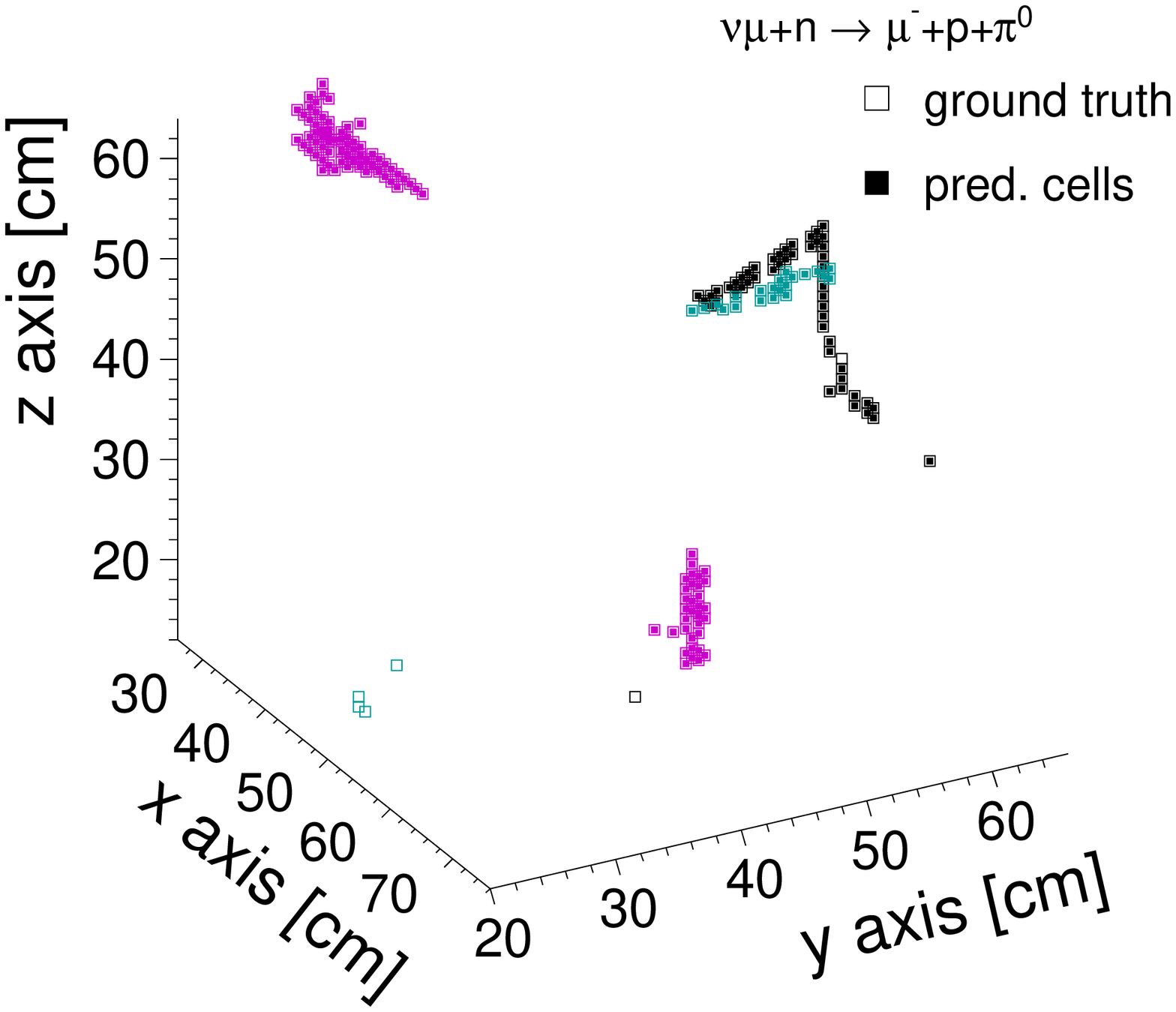}
        \end{center}
    \end{minipage}
\caption{
Plots showing the ground truth of the particle cleanses and labels that are correctly inferred by CNN, where (left) green, cyan, and red are respectively $e^-$, $p$, and $\pi^+$, and (right) black, cyan, and magenta are respectively $\mu^-$, $p$, and $\pi^0$. The training was done using six different neutrino-nucleon interactions (see texts). 
}
\label{fig:5_1_2}
\end{figure}

A left plot of \Figref{fig:5_1_1} shows the hit display of the CC1$\pi$: $e^-+p+\pi^+$ interactions, and a right one shows ground truth. If there are multiple hits in the same pixel, the class that gave the largest light yield is labeled as the ground truth. Although it can be possible to make a class that has a double and triple hit, it has not been done since increasing the number of classes requires more computation time. The current setting of the number of labels can be, however, sufficient to show the performance of the segmentation using CNN. \Figref{fig:5_1_2} show results of the hit segmentation by U-Net where the training was done, as the demonstration, using six different neutrino-nucleon interactions: CCQE, CC1$\pi$: $l^{-}+p+\pi^+$, and CC1$\pi$: $l^{-}+p+\pi^0$ including both of the $\nu_e$ and $\nu_\mu$ processes.   

\Tabref{tab:NeuSgmentation} shows the summary of the segmentation accuracy for each particle. Relatively good accuracies are achieved for the particles that give continuous hits while the accuracies of $n$ and $\pi^+$, which gives out-lands hits by free neutron decay and hadronic shower, are relatively worse, whose segmentation can be also difficult even with the standard clustering algorithms. In particular, it should be noted that the response for $n$ is currently trained using particles from FSI, and the number of samples available for the training is less. Although the number of inputs is limited due to the restriction of the training environment, it can be thought that the accuracies will be improved more by inputting different interactions to give the diversity to the network.

\begin{table}[H]
  \begin{center}
    \caption{
     Segmentation accuracy for each particle after training using six different neutrino-nucleon interactions: CCQE, CC1$\pi$: $l^{-}+p+\pi^+$, and CC1$\pi$: $l^{-}+p+\pi^0$ including both of the $\nu_e$ and $\nu_\mu$ processes. A unit in the table is \%.
}
\vspace{3mm}
    \begin{tabular}{ c | c | c | c | c | c } 
      $e^-$               & $\mu^-$              & $\pi^0$  &
      $\pi^+$            & $n$                     & $p$          \\ \hline \hline
      $93.33\pm0.14$  & $87.35\pm0.22$ & $83.60\pm0.38$ & 
      $78.69\pm0.39$  & $74.71\pm1.23$ & $96.76\pm0.13$
       \\
   \end{tabular}
    \label{tab:NeuSgmentation}
  \end{center}
\end{table}

\subsection{Regression of momentum}

\begin{figure}[htbp]
    \begin{minipage}{0.5\hsize}
        \begin{center}
        \includegraphics[width=76mm]{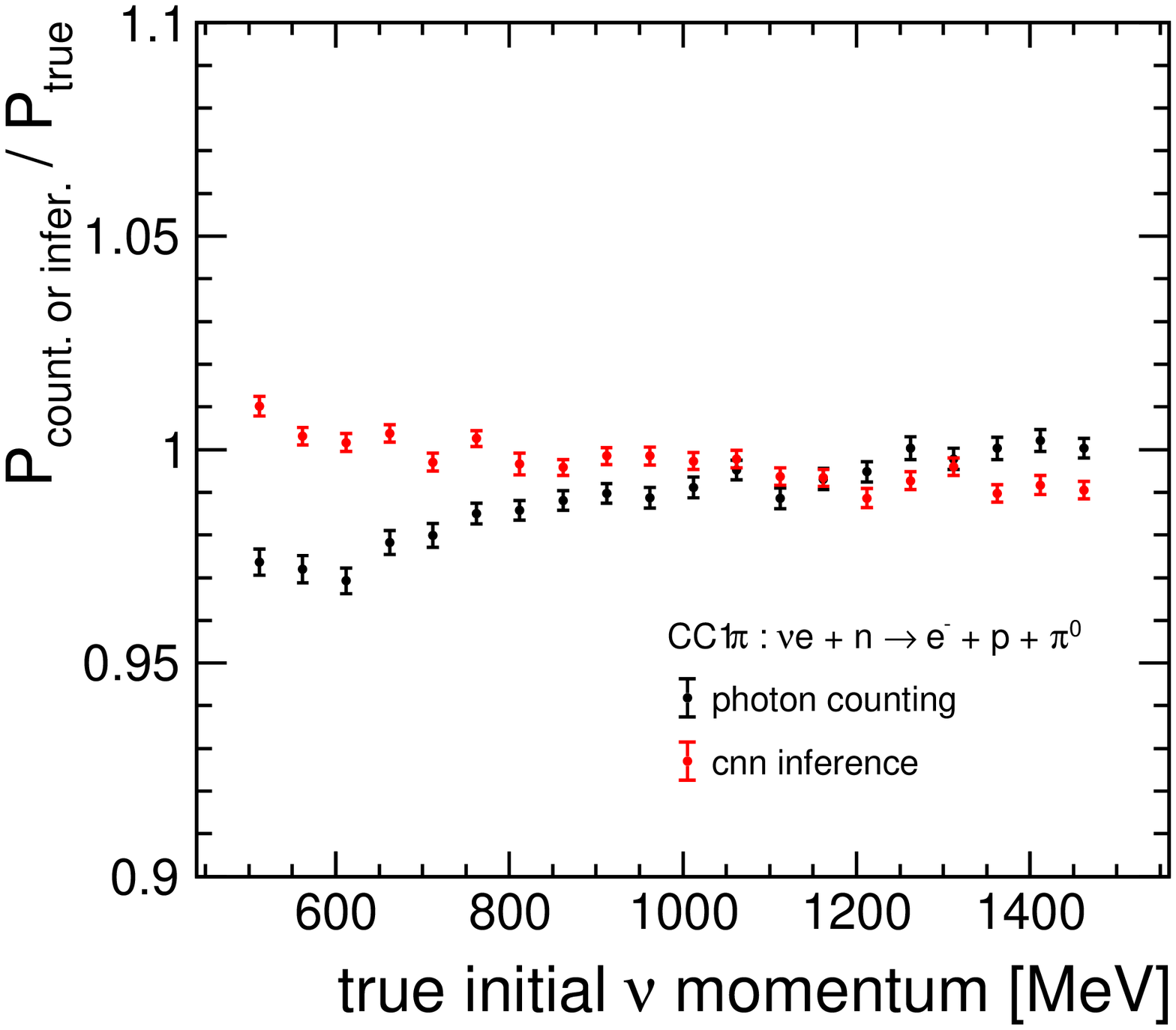}
        \end{center}
    \end{minipage}
    \begin{minipage}{0.5\hsize}
        \begin{center}
        \includegraphics[width=76mm]{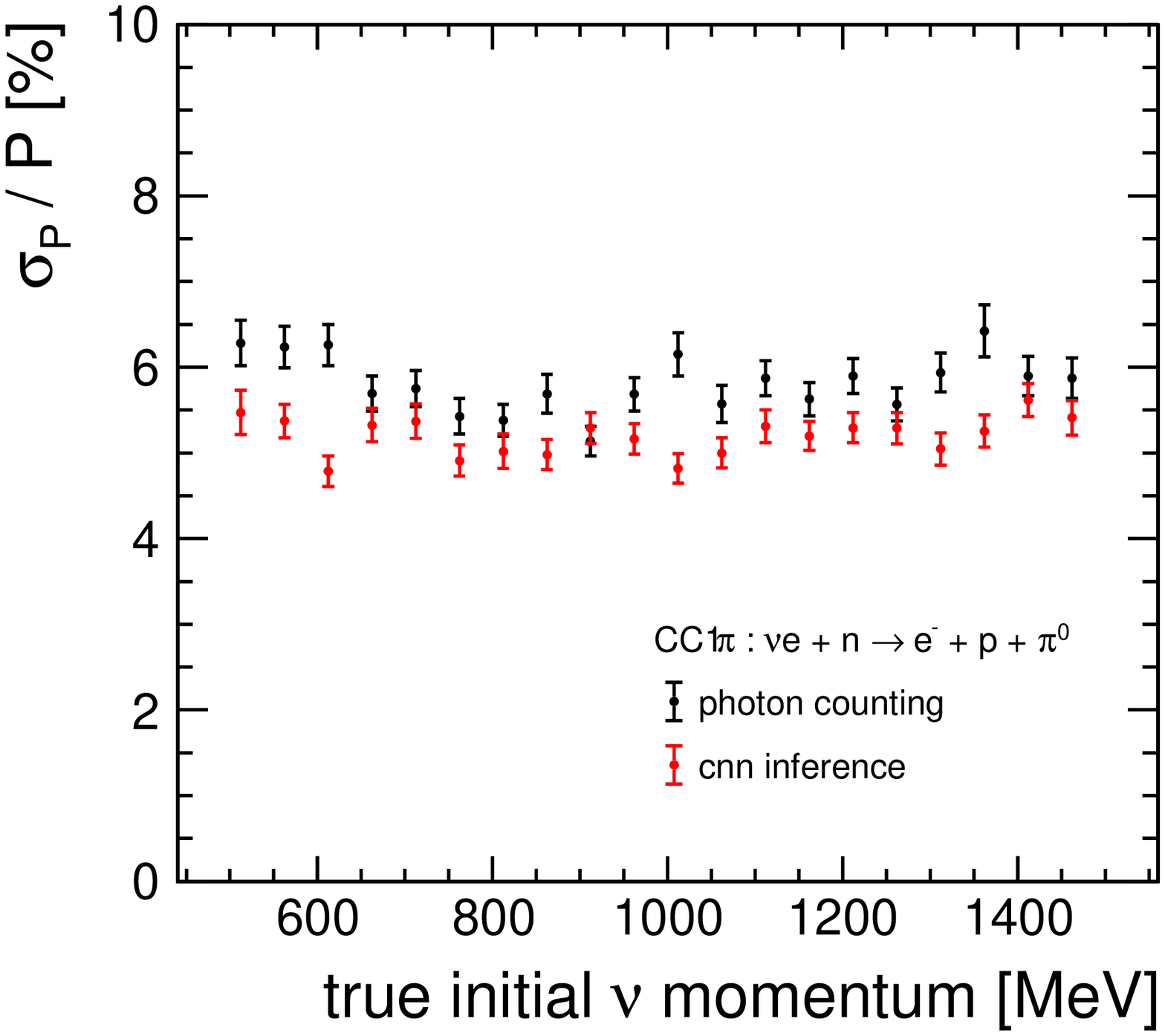}
        \end{center}
    \end{minipage}
\caption{
(Left) The ratio of the reconstructed momenta $P_{\rm count.}$ based on the photon counting and inferred momenta $P_{\rm infer.}$ with CNN to the true momenta $P_{\rm ture}$. (Right) Momentum resolution with both of the methods. The demonstration was done with the CC1$\pi$: $\nu_e+n \rightarrow e^- +p+\pi^0$ process.
}
\label{fig:5_2_1}
\end{figure}

The momentum regression has been also tested for the neutrino-nucleon interactions. \Figref{fig:5_2_1} shows ratio of the reconstructed momentum based on the photon counting and inferred momentum with CNN to the true momentum, where the CC1$\pi$: $\nu_e+p \rightarrow e^- +p+\pi^0$ process is picked up for the demonstration since there exist $e^-$, $\gamma$, and $p$ in the final state. $P_{\rm ture}$ is defined by looking at the energy deposit in the detector since the long radiation length allows a fraction of the particles to escape from the detector volume. Similar to the single particle response, the Birk's suppression is also seen for the photon counting. The momentum resolution by CNN is about 5\%, which is better than that of the photon counting, about 6\%, for the given range. When testing CCQE: $\nu_e+p \rightarrow e^- +p$, the momentum resolution is respectively about 4\% and 5\% with CNN and the photon counting.

\subsection{Classification of neutrino-nucleon events}

\begin{figure}[H]
    \begin{minipage}{0.5\hsize}
        \begin{center}
        \includegraphics[width=76mm]{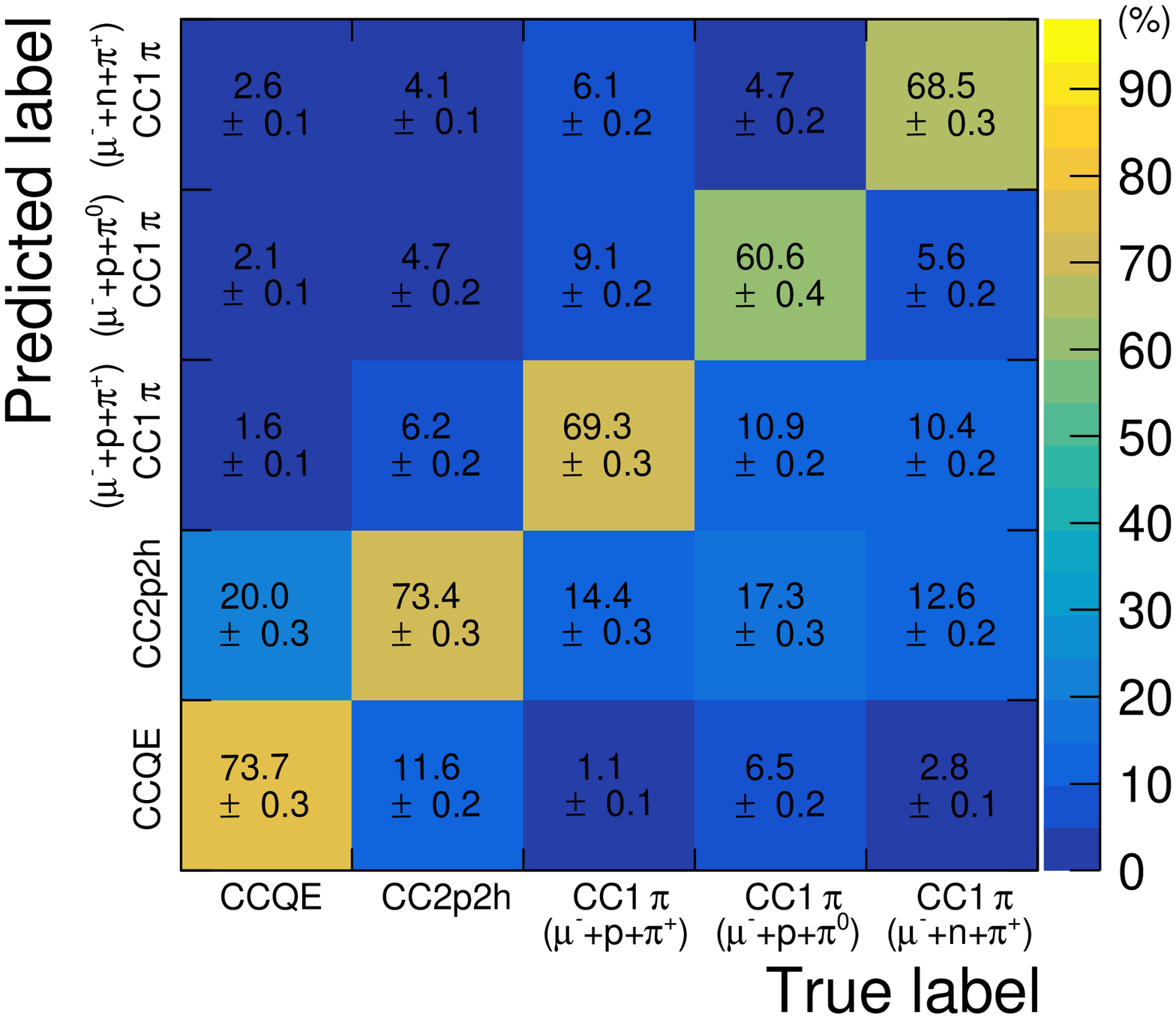}
        \end{center}
    \end{minipage}
    \begin{minipage}{0.5\hsize}
        \begin{center}
        \includegraphics[width=76mm, height=64mm]
        {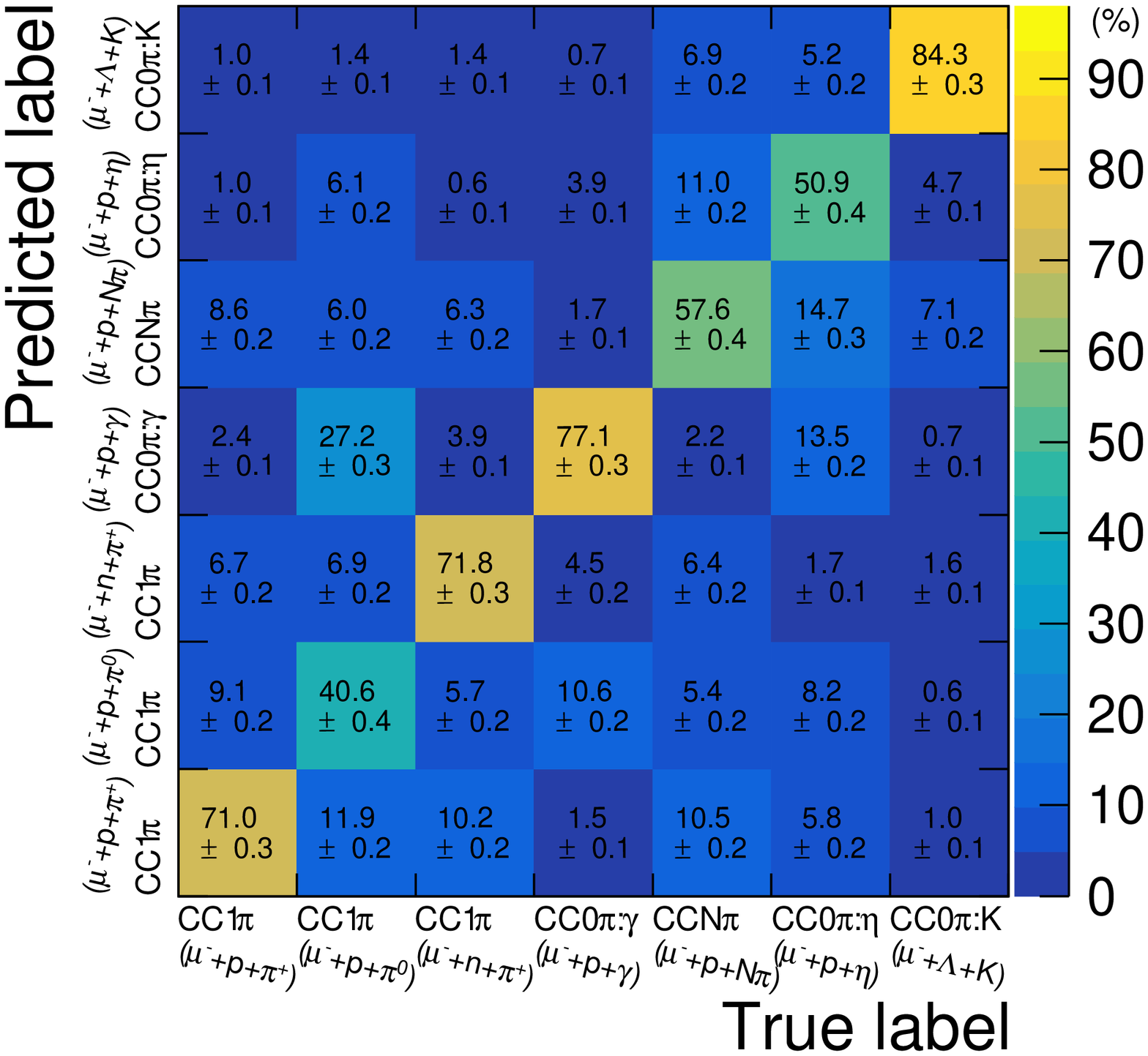}
        \end{center}
    \end{minipage}
\caption{
Confusion matrices for the neutrino-nucleon interactions of $\nu_\mu$: (left) CCQEs vs. CC1$\pi$ and (right) CC1$\pi$ vs. CC0$\pi$. 
}
\label{fig:5_3_1}
\end{figure}

The classification test of neutrino-nucleon interactions was performed in two different patterns: 1). CCQEs vs. CC1$\pi$, and 2). CC1$\pi$ vs. CC0$\pi$. \Figref{fig:5_3_1} show the confusion matrices for 1). and 2) for $\nu_\mu$ (the matrices for $\nu_e$ are shown in appendix). It can be seen that CCQEs and CC1$\pi$ are well separated though some fractions of about 2\% CCQEs exist in CC1$\pi$. 

\begin{figure}[H]
    \begin{minipage}{0.5\hsize}
        \begin{center}
        \includegraphics[width=76mm]{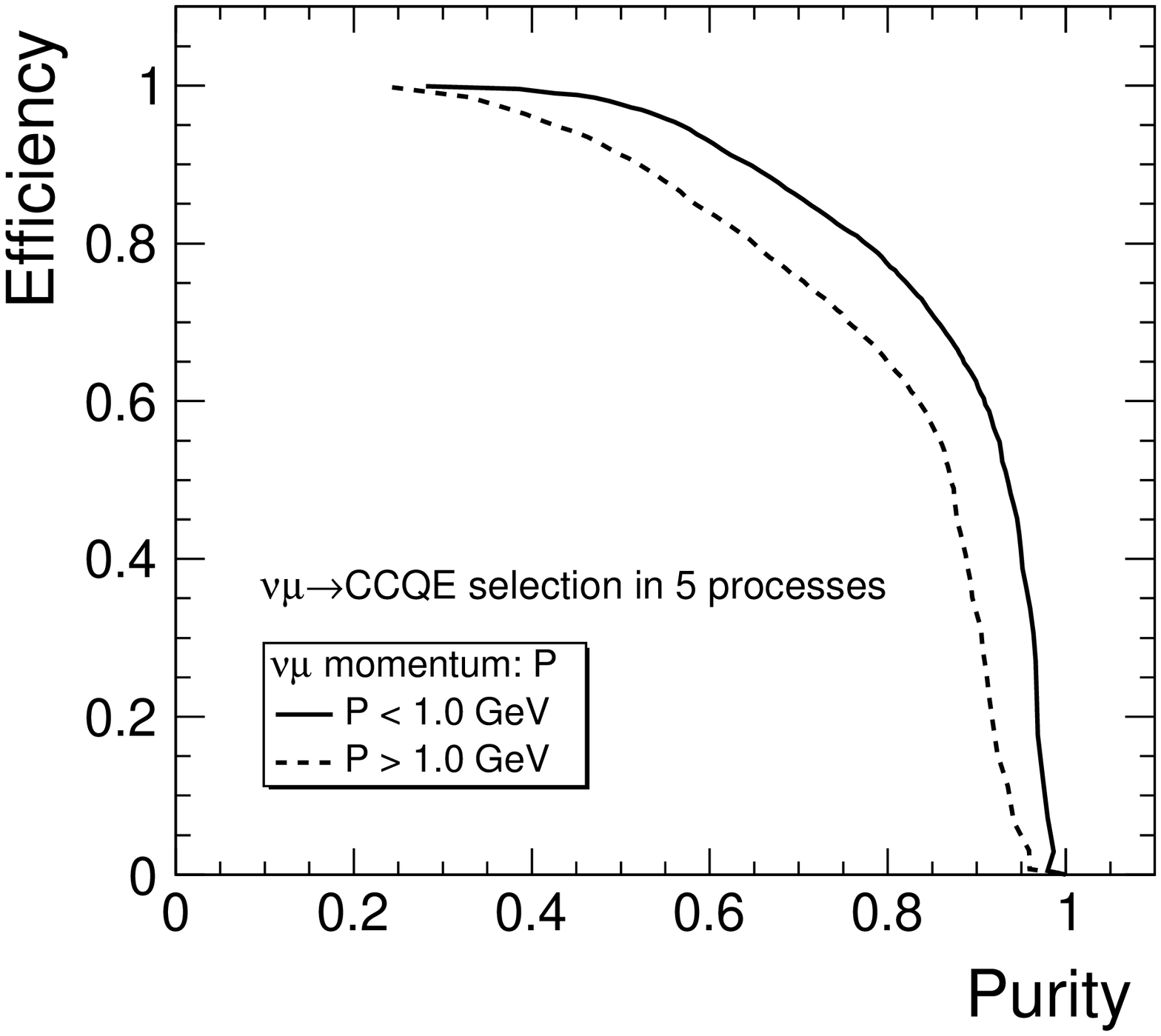}
        \end{center}
    \end{minipage}
    \begin{minipage}{0.5\hsize}
        \begin{center}
        \vspace{+2mm}
        \includegraphics[width=74mm, height=66mm]
        {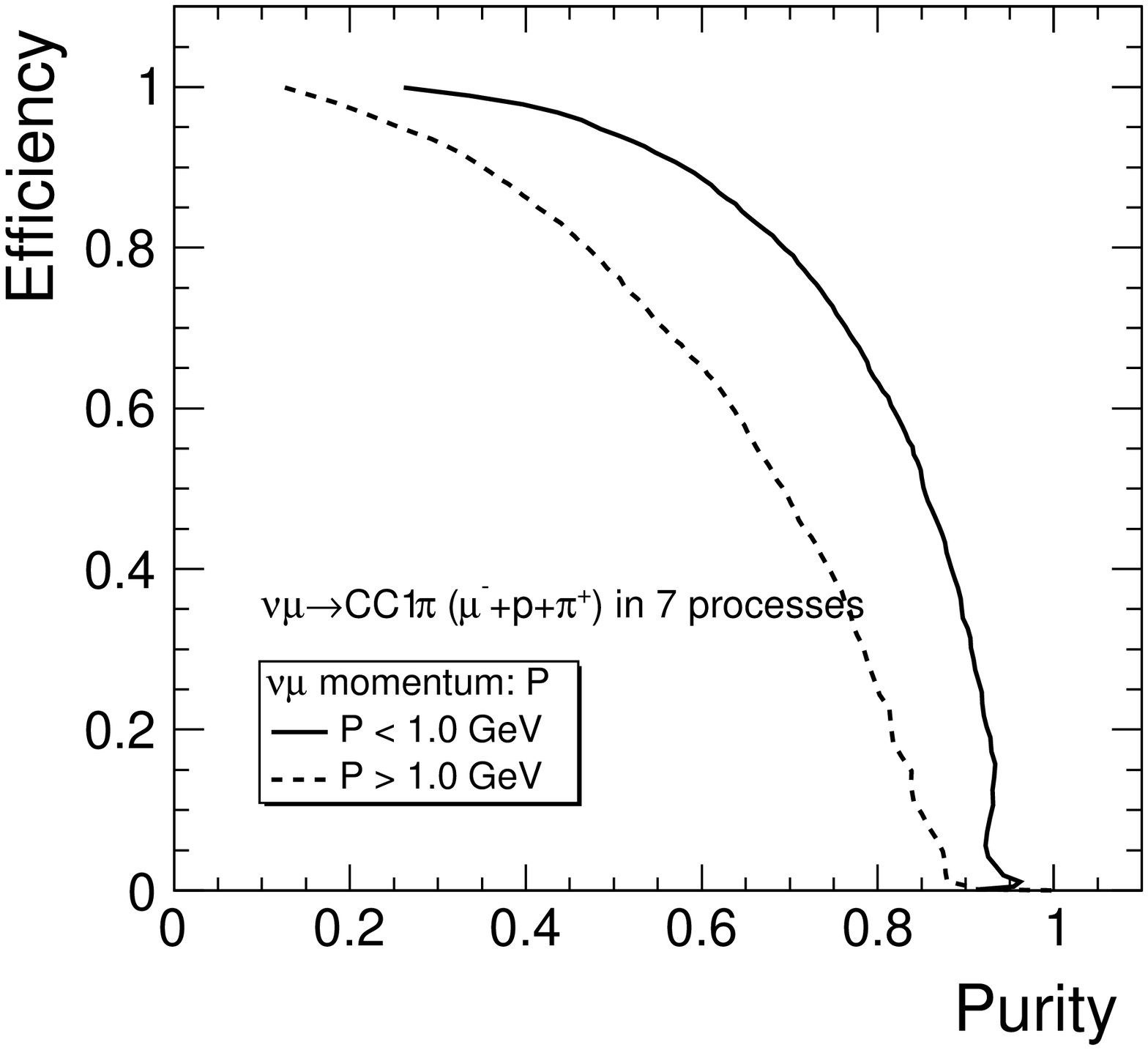}
        \end{center}
    \end{minipage}
\caption{
Plots showing the efficiency vs. purity (background suppression) curves for the $\nu_\mu$ interactions. (Left) CCQE is assumed to be the signal under CCQEs vs. CC1$\pi$. (Right) CC1$\pi$: $\mu^- +p+\pi^+$ is assumed to be the signal under CC1$\pi$ vs. CC0$\pi$.
}
\label{fig:5_3_2}
\end{figure}

When looking at the CCQEs only, the classification accuracy reaches about 73\%. Similarly, when comparing the confusion among CC1$\pi$ vs. CC0$\pi$, the interaction processes involving $\pi^0$ and $\gamma$ in the final state is largely confused. However, as long as focused on the interaction where the charged $\pi$ appears in the final state, the classification accuracy exceeds 70\%. \Figref{fig:5_3_2} show the efficiency vs. purity (background suppression) curves assuming  CCQE and CC1$\pi$: $\mu^- +p+\pi^+$ are the signal process. Depending on the initial $\nu_\mu$ energy, the performance is slightly different though both parameters exceed about 70\%.

Unfortunately, since there are no physics-related studies using the SFGD model, it is not easy to estimate the capability of CNN in terms of the physics performance. Nevertheless, it would be possible to refer to the past results for the evaluation of CNN. In the literature~\cite{PhysRevD.92.112003}, averaged flux-integrated CCQE cross-section of $\nu_\mu$ on a carbon target was measured with an 8\% relative statistical error, under a target mass (fine grained detector 1) of about 0.9 tons, accumulation of $2.6 \times 10^{20}$ protons on target (POT), and a 50 \% data quality cut. An additional 10\% data reduction was made for particle identification and topological cut. The remaining number of events is 5841 with selection efficiency of 40\%, where the CCQE and CC1$\pi$ processes are dominated.

\begin{figure}[htbp]
    \begin{minipage}{0.5\hsize}
        \begin{center}
       \includegraphics[width=76mm]{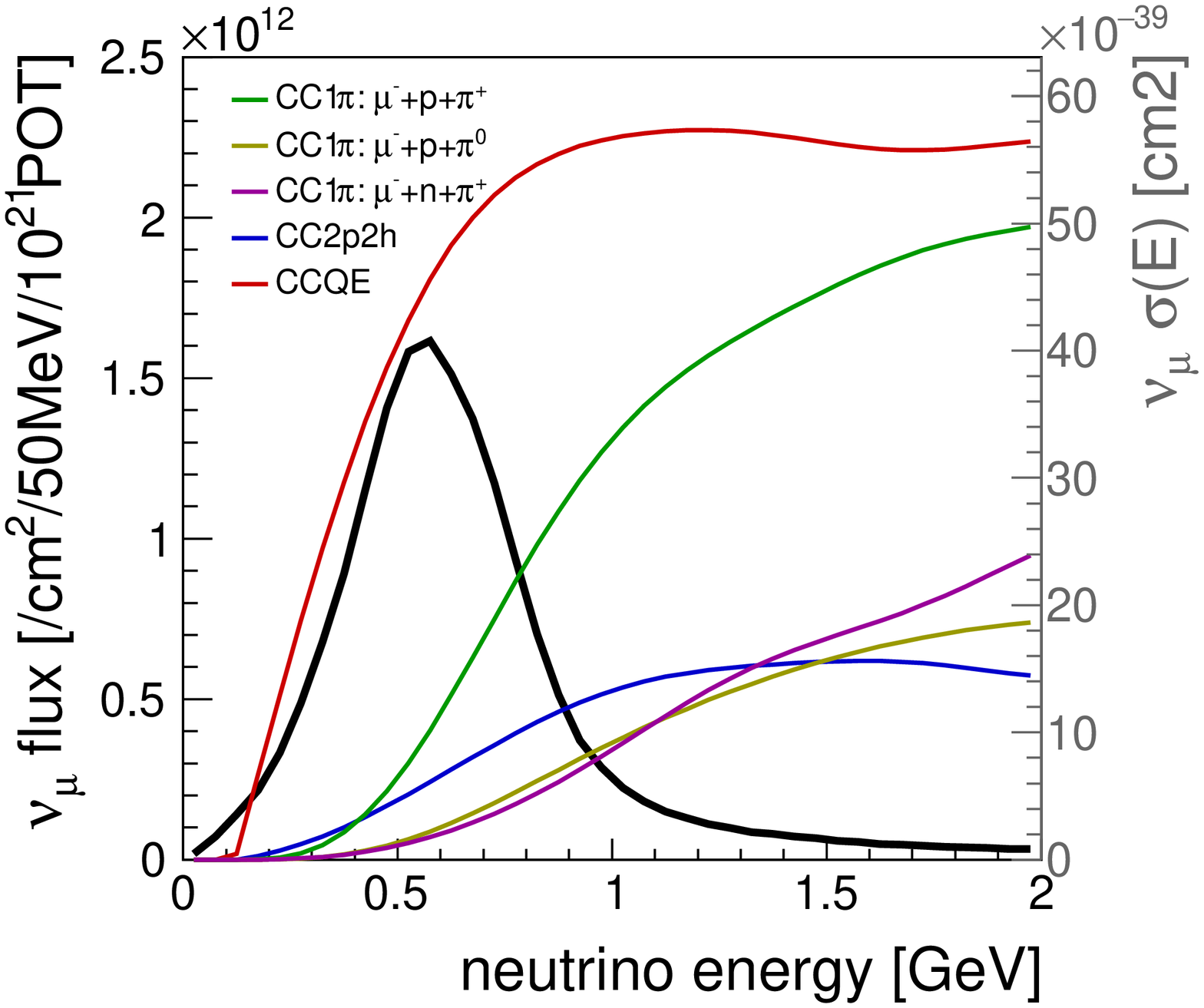}
        \end{center}
    \end{minipage}
    \begin{minipage}{0.5\hsize}
       \begin{center}
       \includegraphics[width=76mm]{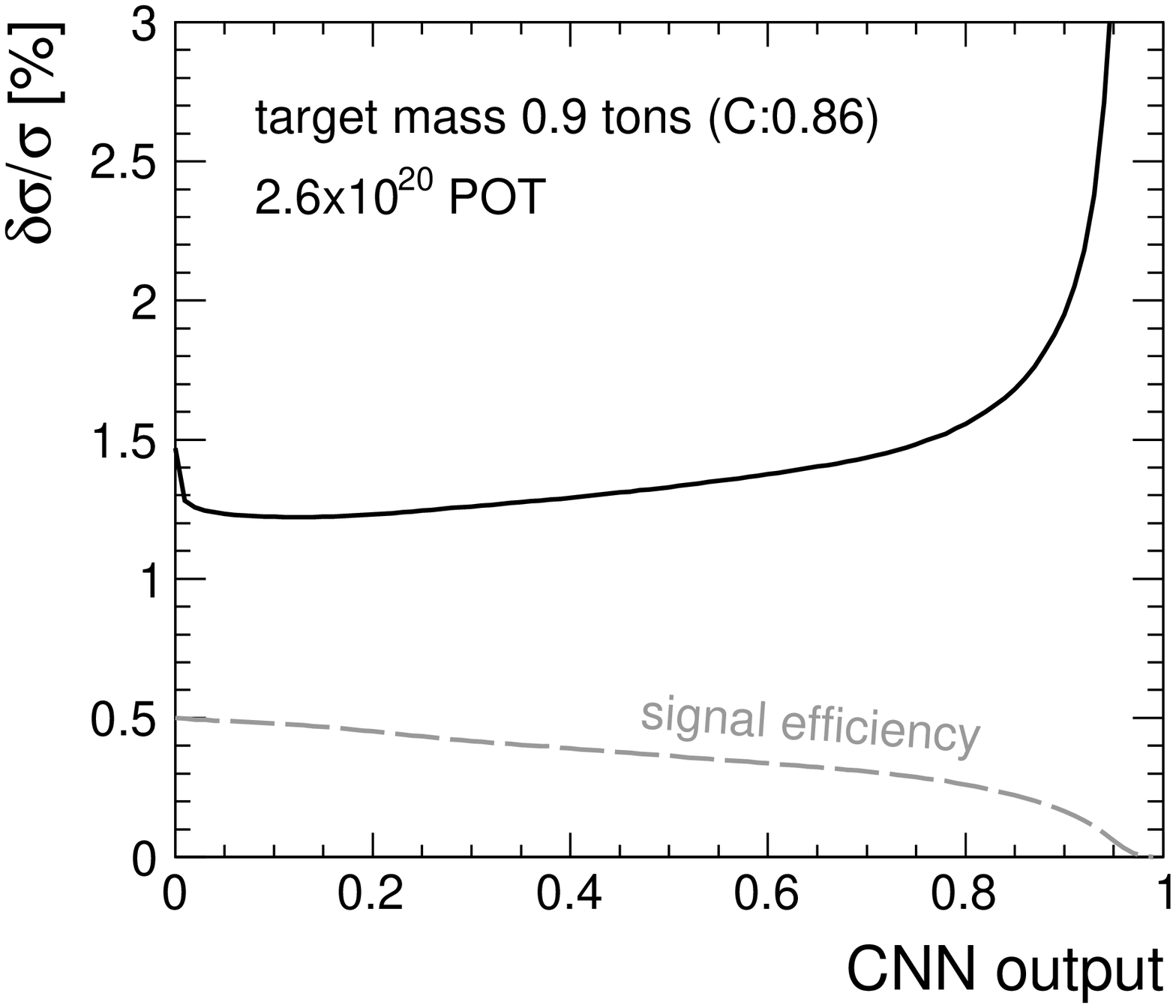}
       \end{center}
    \end{minipage}
	\caption{
(Left) The $\nu_{\mu}$ flux~\cite{berns2021recent} and nominal neutrino-nucleon interaction crosse section of $\nu_\mu$ on the carbon target with NEUT. (Right) The relative statistical error of the flux-integrated CCQE cross-section of $\nu_\mu$ on the carbon target assuming the combination of SFGD and the CNN signal selection.
}
	\label{fig:5_3_4}
\end{figure}

If it had been a combination of SFGD and CNN under the $\nu_{\mu}$ flux~\cite{berns2021recent} and nominal cross-sections with NEUT on the carbon target shown in the left of \Figref{fig:5_3_4}, the relative statistical error would be about 1.3\% at a CNN cut point of 0.12, as shown in the right, where the number of remaining signal and backgrounds are 7636 and 1063 with almost no loss of the signal except the data quality cut.

     


\section{Discussion and summary}
\label{sec:DiscussionSummary}

In the current simulation, electronic noise is not taken into account. But, it is not expected to be a problem since the electronic noise, together with the dark noise of MPPC, is at most a few photons, which is already suppressed in this study. The optical crosstalk in the current simulation is slightly underestimated since the modeling of the actual scintillator cube is quite difficult. However, there exist several reports that adding effects of impurities such as the electronic noise and optical crosstalk into the training can improve the CNN perfromance because it can increase the diversity of the network. Therefore, a few \% variation of the optical crosstalk itself is not expected to degrade the results significantly. In contrast, for actual use in physics analysis, it's necessary to implement a dedicated detector model that can reproduce acquired data well, because the response of the detector greatly affects the training results. In addition, the standard FTFP\_BERT physics model and NEUT parameters are used to generate the events in this study. In fact, response of the energy deposit, development of hadronization, generation of particles via nucleon interactions and so on could vary significantly depending on the physics model set up in GEANT4 and NEUT, which could change the training results as well. Although such tests are beyond the purpose of this study, a good understanding of those systematic uncertainties due to the physics models is necessary for the future physics analysis.

CNN was applied to the response of the single particle and neutrino-nucleon interactions with the scintillator detector subdivided into 1 cubic cm, and the performance of CNN are tested tested for the hit clustering, momentum inference, and event classification tasks. In particular, the inference of the momentum resolution was shown to be improved by a factor of 1.5 for $e^-$ and of 2 for $p$ compared to the conventional photon counting methods. In addition, thanks to the subdivision of 1 cubic cm, the classification accuracy of the single particle and neutrino-nucleon interactions was confirmed to reach about 94\% and 70\% respectively. It is also expected that the training with a sufficient number of event samples can improve the performance of not only the hit clustering but also each task. Using these techniques, it was also shown that the cross-section analyses with CNN is fully feasible.


\clearpage

\appendix
\section*{Classification test of $\nu_e$ interactions}
\begin{figure}[H]
     \begin{minipage}{0.5\hsize}
        \begin{center}
        \includegraphics[width=76mm]{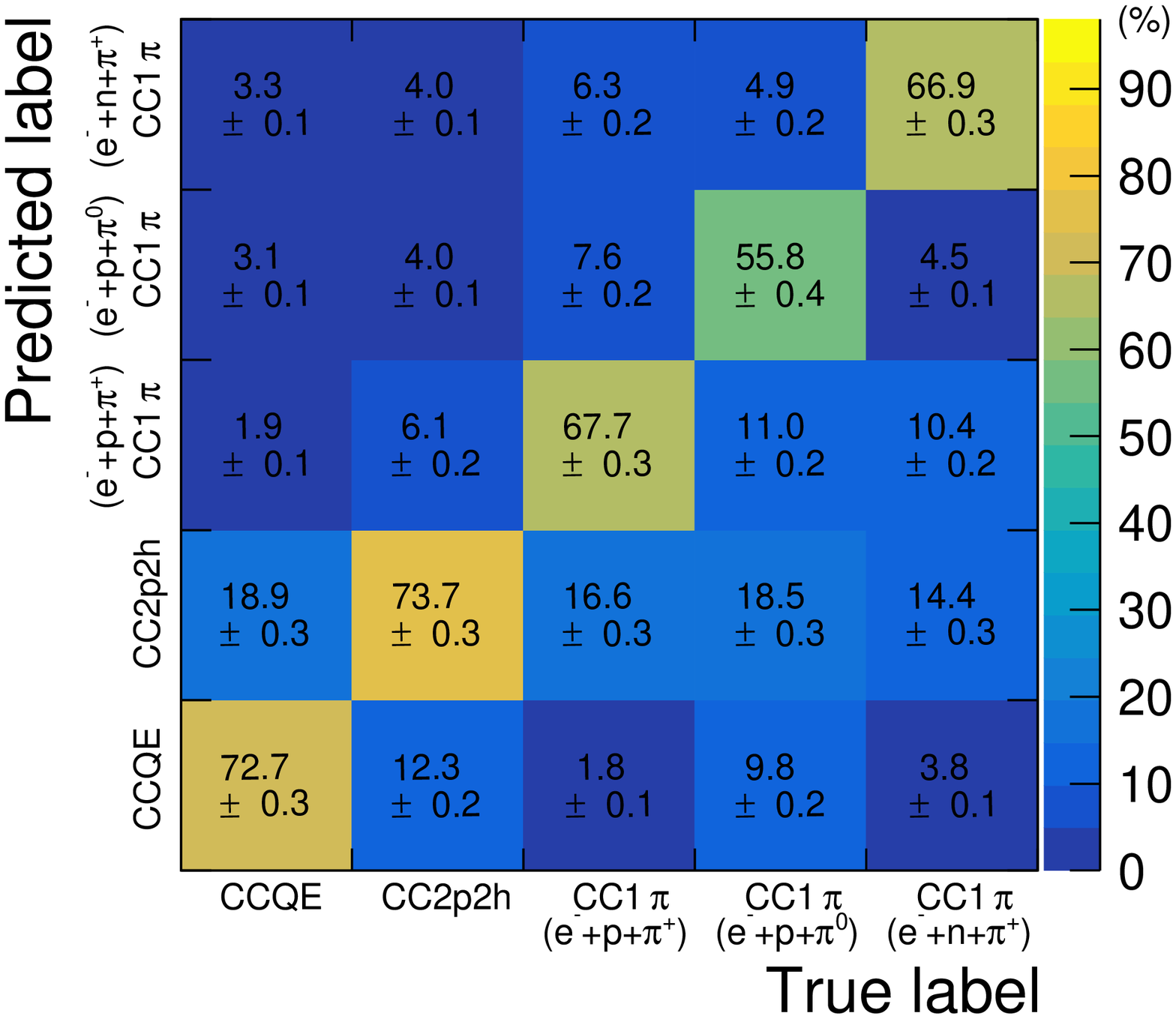}
        \end{center}
    \end{minipage}
    \begin{minipage}{0.5\hsize}
        \begin{center}
        \includegraphics[width=76mm, height=64mm]
        {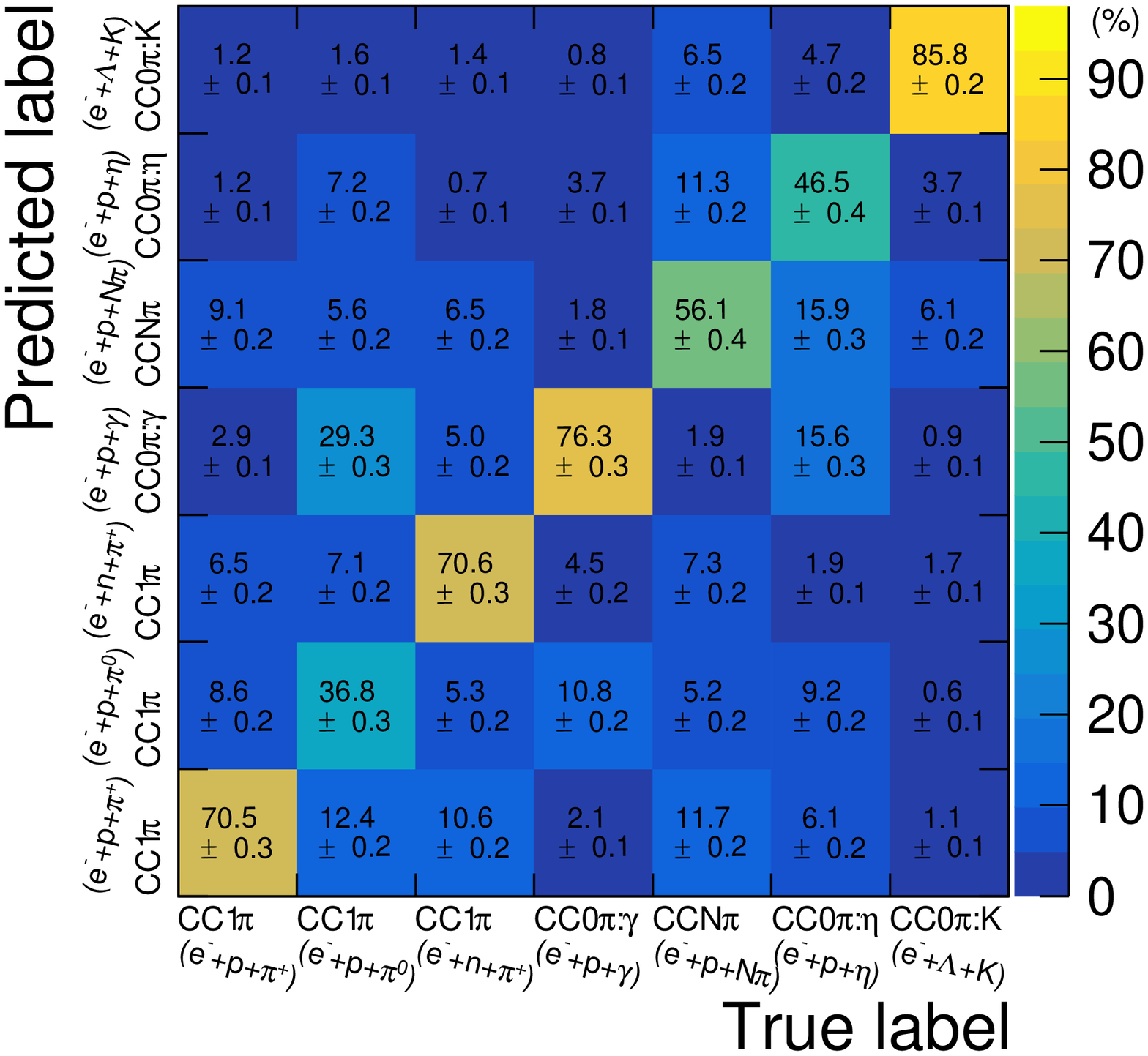}
        \end{center}
    \end{minipage}
\caption{
Confusion matrices for the neutrino-nucleon interactions of $\nu_e$: (left) CCQEs vs. CC1$\pi$ and (right) CC1$\pi$ vs. CC0$\pi$. 
}
\label{fig:Append1}
\vspace{5mm}
%
    \begin{minipage}{0.5\hsize}
        \begin{center}
        \includegraphics[width=76mm]{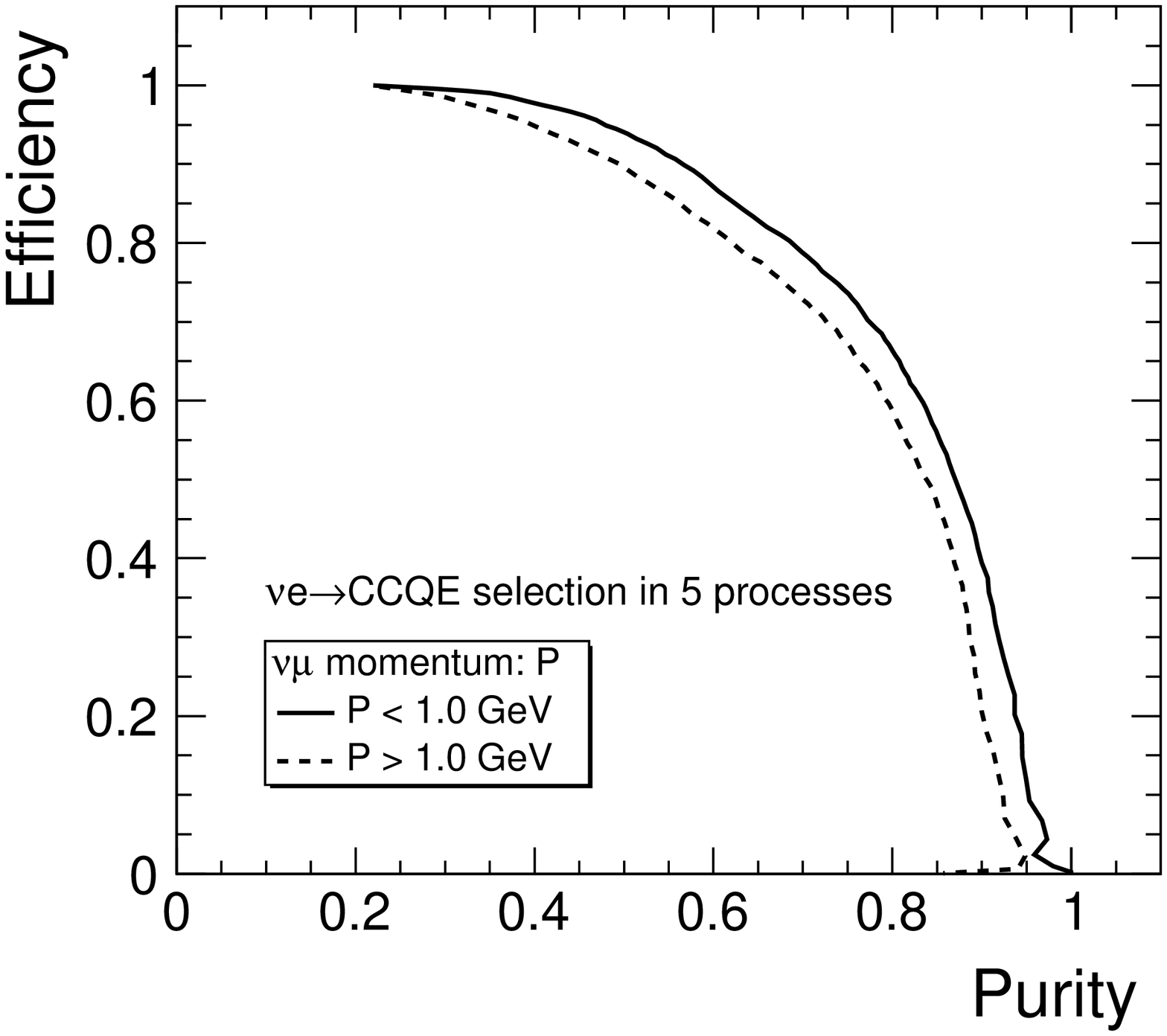}
        \end{center}
    \end{minipage}
    \begin{minipage}{0.5\hsize}
        \begin{center}
       \vspace{+2mm}
        \includegraphics[width=74mm, height=66mm]
         {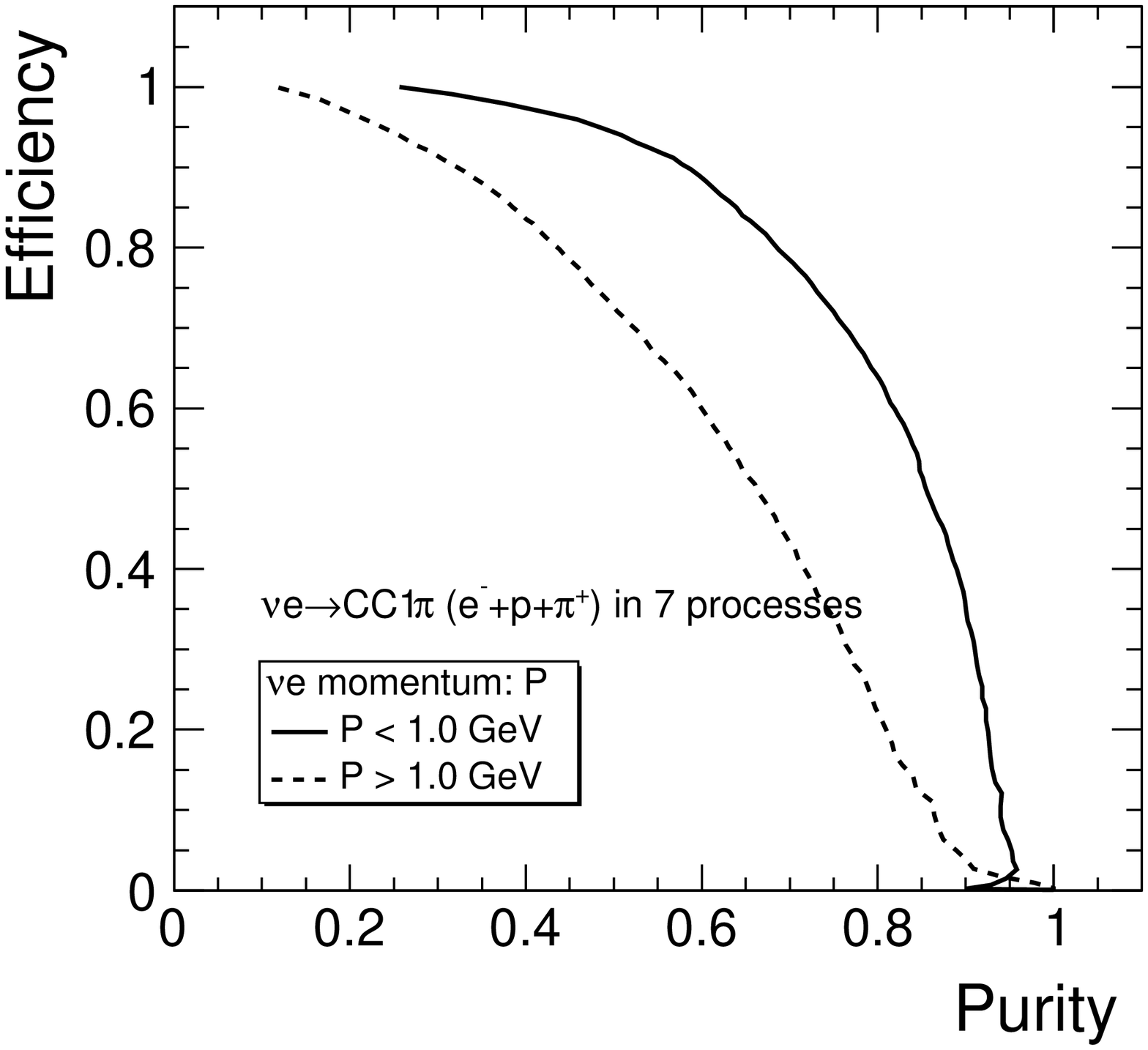}
        \end{center}
    \end{minipage}
\caption{
Plots showing the efficiency vs. purity (background suppression) curves for the $\nu_e$ interactions. (Left) CCQE is assumed to be the signal under CCQEs vs. CC1$\pi$. (Right) CC1$\pi$: $e^- +p+\pi^+$ is assumed to be the signal under CC1$\pi$ vs. CC0$\pi$.
}
\label{fig:Append2}
\end{figure}

The classification test of neutrino-nucleon interactions for 1). CCQEs vs. CC1$\pi$, and 2). CC1$\pi$ vs. CC0$\pi$. \Figref{fig:Append1} show the confusion matrices for 1). and 2) for $\nu_e$. \Figref{fig:Append2} show the efficiency and purity curves for each.

\clearpage

\bibliographystyle{JHEP}
\bibliography{references}

%



\end{document}